\documentclass[twocolumn]{aastex62}
\usepackage{color} %https://www.sharelatex.com/learn/Using_colours_in_LaTeX
\usepackage{lipsum}
\usepackage{mathtools}
\usepackage{placeins}
\usepackage{multirow}
\usepackage{float}
\usepackage{amsmath}
\usepackage{physics}
\usepackage{silence}
\usepackage{multirow, booktabs}
\newcommand{\shadowCIB}{final_shadow_CIB_importance_ISM_dust_sync_free}
\newcommand{\shadowISMdust}{final_shadow_ISM_dust_importance_sync_free_CIB}
\newcommand{\MCMCwithISMdustPC}{final_all_foregrounds_with_ISM_dust_4_PC}

\shorttitle{Impact of Dust on CMB Spectral Distortions Measurements}
\shortauthors{Zelko and Finkbeiner}

\begin{document}
\title{Impact of Dust on Spectral Distortion Measurements of the Cosmic Microwave Background}

\author{Ioana A. Zelko}
\affiliation{Harvard-Smithsonian Center for Astrophysics, 
60 Garden Street,
Cambridge, MA  02138, USA}
\author{Douglas P. Finkbeiner} 
\affiliation{Harvard-Smithsonian Center for Astrophysics,
60 Garden Street,
Cambridge, MA 02138, USA}
\affiliation{Department of Physics, Harvard University,
17 Oxford Street,
Cambridge, MA 02138, USA}
%\tableofcontents

\begin{abstract}
Spectral distortions of the cosmic microwave background (CMB) are sensitive to energy injection by exotic physics in the early universe.  The proposed Primordial Inflation Explorer (PIXIE) mission has the raw sensitivity to provide meaningful limits on new physics, but only if foreground emission can be adequately modeled.  We quantify the impact of interstellar dust on Compton $y$ and $\mu$ measurements by considering a range of grain size distributions and compositions constrained by theoretical and observational priors \citep{Zelko2020}.  We find that PIXIE can marginalize over a modest number of dust parameters and still recover $y$ and $\mu$ estimates, though with increased uncertainty.  As more foreground components are included (synchrotron, free-free), the estimates of $y$ degrade, and measurement of $\mu$ in the range sometimes considered for the standard $\Lambda$CDM of $2\times10^{-8}$ becomes infeasible without ancillary low-frequency foreground information. An additional concern is dust absorption of the CMB monopole, a subtle effect that must be included.  We quantify one form of model discrepancy error, finding that the error introduced by fitting our interstellar medium dust model with a modified blackbody is too large for CMB spectral distortions to be detectable. 
The greatest challenge may be the cosmic infrared background (CIB).  We find that $\mu$ and $y$ are extremely sensitive to modeling choices for the CIB, and quantify biases expected for a range of assumptions. 
\end{abstract}


\section{Introduction}\label{sec:intro}
Over the past 30 years, the cosmic microwave background (CMB) has profoundly influenced our understanding of the history of our universe, and ushered in an era of precision cosmology.  Spatial anisotropy (both polarized and unpolarized) has been the focus of both experiment and theory, and has driven the field forward. However, spectral distortions in the CMB have never been detected. 

\subsection{Spectral Distortions}
Many physical processes can lead to departures from the blackbody spectrum of the cosmic radiation \citep{Chluba2014,Chluba2014a,Tashiro2014,Hill2015}, for example: reionization and structure formation, decaying or annihilating particles, dissipation of primordial density fluctuations, cosmic strings, primordial black holes, small-scale magnetic fields, adiabatic cooling of matter, and recombination.

The distortions created between double Compton scattering decoupling at $z \approx 10^6$ and the decoupling of thermalization by Compton scattering at $z \approx 10^5$ are usually characterized by a chemical potential and are thus called $\mu$ distortions \citep{Zeldovich1969, Sunyaev1970a, Illarionov1975, Burigana1991,Hu1993a}. After thermalization decoupling, inverse Compton scattering or other similar processes produce what we call $y$-type distortions \citep{Zeldovich1969, Sunyaev1972}. Energy injection at intermediate redshifts ($1.5\times 10^4 < z < 2\times 10^5$) produces a distortion spectrum that is intermediate between $y$ and $\mu$ distortions \citep{Chluba2012, Chluba2014a,Khatri2012}.

The FIRAS experiment on NASA's Cosmic Background Explorer (COBE) \citep{Boggess1992} provided upper limits on the $\mu$ and $y$ parameters at values of $|y| < 1.5 \times 10^{-5} , |\mu|  < 9 \times 10^{-5} $. It put constraints on the deviation of the observed spectrum from the blackbody spectrum of
$\frac{\Delta I_{\nu}}{I_{\nu}} \leq 10^{-5} - 10^{-4}$ \citep{Mather1994,Fixsen1996,Fixsen2009}.  Although these constraints were obtained with hardware designed nearly 40 years ago, they have never been superseded.  With modern technology, the sensitivity to spectral distortions could be improved by a few orders of magnitude. 

\subsection{The Primordial Inflation Explorer}
The Primordial Inflation Explorer, PIXIE  \citep{Kogut2011,Kogut2016,Naess2019,Kogut2020}, is a proposed satellite mission that aims to map the absolute intensity and linear polarization (Stokes $I$, $Q$, and $U$ parameters) of the CMB. It would use 416 spectral channels from 14.4 GHz to 6 THz to map the whole sky with an angular resolution of $2.6^{\circ}$.  Like its predecessor, FIRAS \citep{Mather1994,Fixsen1996,Fixsen2009}, PIXIE would use a polarizing Michelson interferometer with a Fourier transform spectrometer, but with 76 times greater sensitivity. 

This increased sensitivity would both improve spectral distortion constraints and also constrain primordial gravity waves from the inflationary epoch to $r < 10^{-3}$ at more than 5$\sigma$ based on the CMB anisotropy.

PIXIE is very sensitive to the thermal radiation from dust, and will provide important information about the interstellar medium (ISM). However, the dust foreground poses a challenge for the detectability of spectral distortions. The sensitivity of PIXIE to spectral distortions depends on how well the dust emission can be modeled, which in turn depends on both the correctness of the dust model and its degeneracy with $y$ and $\mu$. In the FIRAS era, the dust spectral energy distribution (SED) was represented by a modified blackbody (MBB) \citep{Reach1995} or a combination of MBBs \citep{Finkbeiner1999}. This was used and improved by the Wilkinson Microwave Anisotropy Probe (WMAP) team \citep{Bennett2003}, and by \emph{Planck} \citep{PlanckCollaboration2014, Collaboration2016a}. \cite{Abitbol2017} studied the overall impact of foregrounds while using this dust model, and mentioned the need for an analysis using a more comprehensive dust model.  \cite{Chluba2017,Remazeilles2020}, and \cite{Rotti2021} considered how using a moment expansion approach together with the internal linear combination (ILC) may provide a way to make use of the spatial information for foreground subtraction for extracting average-sky signals and CMB anisotropies, as an alternative to exactly modeling foregrounds. At this time, the various models describing temperature distribution, the abundance and composition of the dust, and the magnetic alignment of aspherical grains are unconstrained. This requires  a broader look at  the possible size distribution of dust grains beyond the modified blackbody dust model. Our goal is to provide a better estimation of the impact dust can have on the PIXIE spectral distortion measurement by studying the impact that varying the dust composition, size distribution, and interstellar radiation field (ISRF) can have on the model. 

\subsection{Modeling Dust}
Of primary concern is thermal emission from dust, both in the Milky Way and in distant galaxies. The size and composition of dust grains are variable from place to place in our Galaxy, and possibly across cosmic time.  We have ideas about the chief constituents of dust, but detailed knowledge is elusive.  Nevertheless, the thermal emission from a wide range of possible dust grains is somewhat similar, and it is plausible that the variability of the emission spectrum can be expressed by a few parameters.  

The primary goal of this work is to quantify the variability of dust emission in the PIXIE frequency range, and estimate the sensitivity of PIXIE to $y$ and $\mu$ distortions after marginalizing over dust emission.  In order for this estimate to be meaningful, it is \emph{not} required that the dust model be correct, merely that it reflect the range of possible variations in the size distribution and composition for interstellar dust.  A model with reasonable constituents and a plausible range of grain size distributions, constrained by well-motivated physical priors, provides at least a \emph{lower bound} on the variation in the emission spectrum.   

\subsection{Summary of Results}
This work presents three main results. First, we start from the parameters describing the size distribution of the grains of dust, which have been constrained to match the existing extinction law variability by \cite{Zelko2020}. The goal is to determine the feasibility of dust foreground subtraction for PIXIE. By exploring the space that the size distributions can span while still maintaining what we know about the dust reddening curve and its spatial variation, we have the opportunity to asses the impact of a broad class of dust models on PIXIE sensitivity. We can also explore the effect of having dust grains exposed to different ISRFs. 

For each sample from the size distribution parameter space, we calculate the dust emission at 416 frequencies to be observed by PIXIE.  Each of these spectra can be thought of as a point in a 416-dimensional vector space.  This ensemble of points occupies a low-dimensional subspace, which we can explore by computing the  principal components (PCs) in this space.  We then use the principal components to perform a Markov Chain Monte Carlo (MCMC) and Fisher information matrix analysis to determine the impact this improved dust modeling will have on the detectability of $y$ and $\mu$ distortions. Our foreground model includes synchrotron radiation, cosmic infrared background (CIB), and free-free emission. We also include the distortions coming from the measurement of the temperature of the blackbody to the required precision. The effect of using the incorrect model to describe the ISM dust is also characterized.

Second, this work explores the result of performing a CIB fit to the contributions coming from dust in galaxies at different temperatures. We create a mock CIB SED, from a superposition of modified blackbodies at different temperatures.  We compute the biases that result from fitting the CIB with a simple MBB, a smoothed MBB, and a PC analysis (PCA). 

Finally, we study the effect of neglecting the absorption of the CMB's monopole component by the  interstellar dust and by dust in other galaxies on the detectability of the spectral distortions. \cite{Nashimoto2020} studied the impact of the ``CMB shadow'' caused by galactic matter on the measurements of CMB polarization and temperature anisotropy. Prompted by their work, we use an MCMC to explore the offset introduced by failing to model this effect both in the ISM dust and the CIB, and the corresponding deviations in the spectral distortion parameters.

In \S \ref{sec:modeling} we explain the modeling of spectral distortions, the other foregrounds, and the PIXIE mission configuration. \S \ref{sec:dust_modeling} describes the modeling of the ISM dust from the dust properties constrained by \cite{Zelko2020} and the PCA method applied to characterize the ISM dust emission. In \S \ref{sec:ISM_dust_forecasting_methods} we characterize the impact of the ISM dust modeling using both the Fisher information matrix method and an MCMC for foreground analysis. The results of using an incorrect dust model are also explored. \S \ref{sec:CIB} shows the analysis of a mock CIB created from superposition of multiple temperature sources. \S \ref{sec:CMB_shadows} presents the analysis and results from including the CMB shadows given by the ISM dust and the dust in other galaxies as part of the model. Finally, we conclude in \S \ref{sec:conclusions}.

\begin{figure*}[t]
	\includegraphics[scale=1]{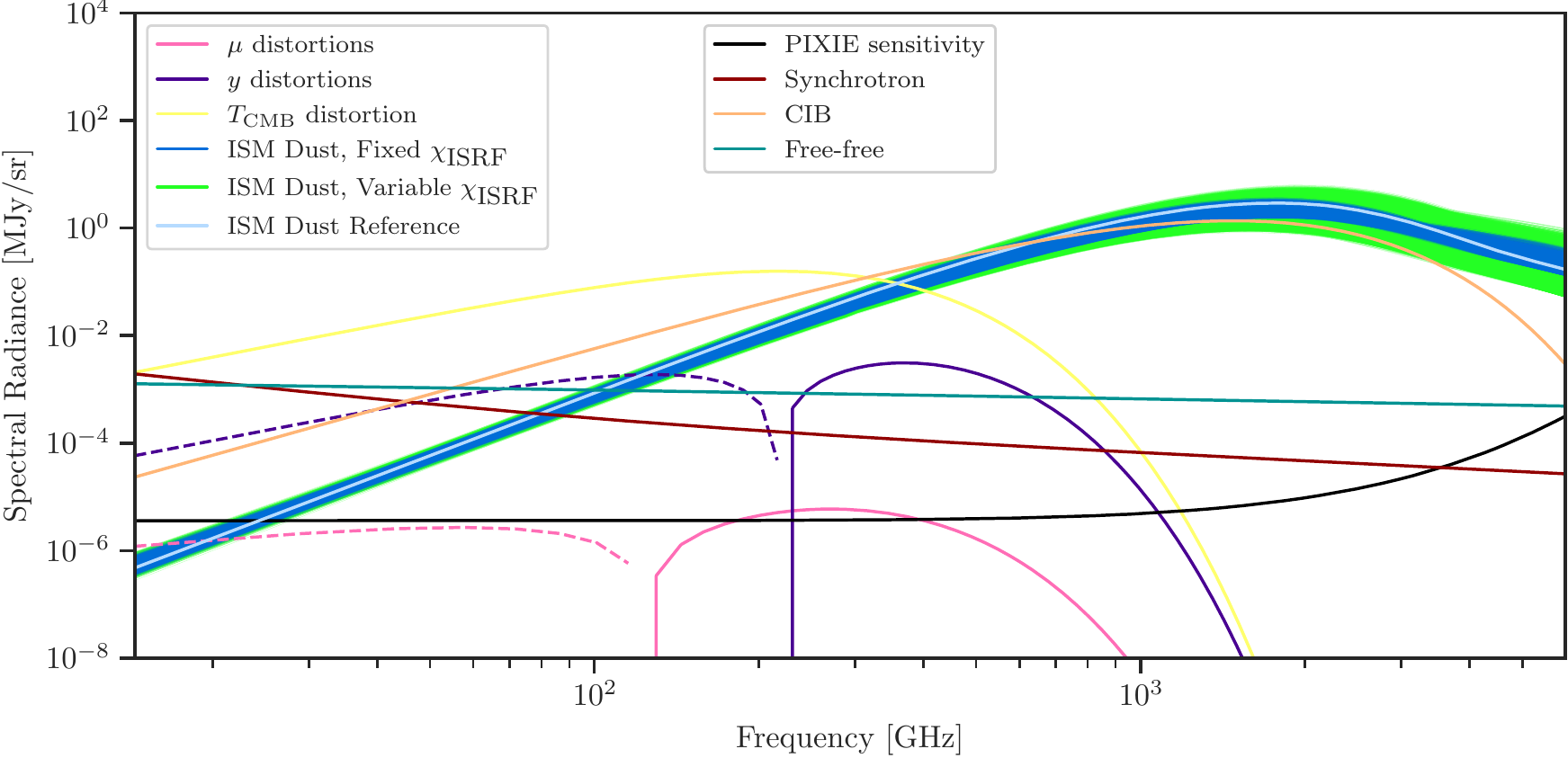}
	\caption{Spectral distortions and various foregrounds with intensities typical at high Galactic latitude. Interstellar dust is modeled with Draine-like models with grain size distributions constrained by optical/UV extinction, far IR emission, and interstellar abundances \citep{Zelko2020}, for a fixed interstellar radiation field (blue) and variable radiation field (green) with the coefficient $\chi_{\textrm{ISRF}}$ taking values between 0.5 and 2.  The dust reference line (light blue) is a sample SED from \cite{Zelko2020} used to generate fiducial values. Negative values are represented by a dashed line. \label{fig:foreground}}
\end{figure*}
\section{Spectral Distortions, Foreground, and Mission Modeling} \label{sec:modeling}
%%%%%%%%%%%%%%%%%%%%%%%%%%%%%%%%%%%%%%%%%%%%%%%%%%%%%%%%%%%%%%%%%%%%%%%%%
%%%%%%%%%%%%%%%%%%%%%%%%%%%%%%%%%%%%%%%%%%%%%%%%%%%%%%%%%%%%%%%%%%%%%%%%%

This work explores the detectability of CMB spectral distortions in the presence of other foregrounds, with particular emphasis on the potential impact of interstellar dust, both in the Milky Way and in other galaxies that produce the CIB. 

For the CMB spectral distortions we consider only the $y$ and $\mu$ distortions, as presented in \S \ref{sec:spectral_distortions_modeling}.
We consider two other foreground components in our analysis: synchrotron radiation and bremsstrahlung (free-free) emission, as described in \S \ref{sec:foreground_modeling}. Reference foregrounds and spectral distortions are plotted in Figure \ref{fig:foreground}, and the fiducial values for the parameters in the model are given in Table \ref{table:foreground_fiducial_table}. 
%%%%%%%%%%%%%%%%%%%%%%%%%%%%%%%%%%%%%%%%%%%%%%%%%%%%%%%%%%%%%%%%%%%%%%%%%%%
%%%%%%%%%%%%%%%%%%%%%%%%%%%%%%%%%%%%%%%%%%%%%%%%%%%%%%%%%%%%%%%%%%%%%%%%%%%
\subsection{Spectral Distortion Modeling}\label{sec:spectral_distortions_modeling}

\begin{table*}[t]
\centering
\begin{tabular}{rrrrrrrrrrr}
\hline
Parameter Name  & $y$ & $\mu$& $\Delta_T$[K] & $ A_{\textnormal{s}}$[MJy/sr] & $\alpha_{\textnormal{s}} $&$\omega_\textnormal{s}$ & $\alpha_{\textnormal{CIB}}$[MJy/sr] & $\beta_{\textnormal{CIB}} $ & $T_{\textnormal{CIB}}$[K] & $A_{\textnormal{FF}}$[MJy/sr]\\ \hline
Fiducial Values & 1.7e-06&2.0e-08&1.2e-04&2.9e-04&-0.82&0.20&1.779e-03&0.86&18.80&3.0e-04 \\ \hline
\end{tabular}
\caption{The fiducial values of the spectral distortions and foreground parameters. The fiducial values for the ISM dust parameters are discussed in \S \ref{sec:dust_modeling} and \S \ref{sec:ISM_dust_forecasting_methods}. \label{table:foreground_fiducial_table}}
\end{table*}

\paragraph{y Distortion}

PIXIE can shed light on the history of star formation by looking at the spectral distortions produced during the epoch of reionization. The CMB photons inverse Compton scatter off of gas that has been ionized by early stars, and generally acquire a small amount of energy from each scattering. These Compton distortions are parameterized by a parameter, $y$, proportional to the energy gain per scattering and the probability of scattering. In practice, it is proportional to the ionized gas pressure, integrated along the line of sight. 

The intensity contribution is given by:

\begin{equation}\label{eq:y}
\Delta I_{\nu}^y = I_0 \frac{x^4e^x}{(e^x-1)^2}\bigg[x \coth\bigg(\frac{x}{2}\bigg) -4 \bigg] y,
\end{equation}
where $I_0 = \frac{2h}{c^2}(\frac{kT_0}{h})^3 = 270$~MJy/sr  for $T_0 = 2.72548$K, and $x = \frac{h \nu}{k T_0}$.

\cite{Hill2015} estimated that reionization, galaxy groups and clusters, and the intracluster medium will create a total signal of $y=1.77 \times 10^{-6}$. Following \cite{Abitbol2017}, we take $y= 1.7 \times 10^{-6}$ as our fiducial value (Figure \ref{fig:foreground}). The spectrum of the $y$ distortion is negative below some frequency (a deficit of photons that have been upscattered) and positive above it.  The zero crossing of Eq. (\ref{eq:y}) is the solution of $ x \coth(\frac{x}{2}) =4 $, which can also be written as $\exp(x) = \frac{4+x}{4-x}$. Its positive solution is $x\approx 3.83$, which corresponds to a frequency of $217.5$~GHz.
%%%%%%%%%%%%%%%%%%%%%%%%%%%%%%%%%%%%%%%%%%%%%%%%%%%%%%%%%%%%%%%%%%%%%%%%%%%
\paragraph{$\mu$ Distortion}

The amplitude of density fluctuations during inflation translates into energy injection in the CMB \citep{Chluba2019} for redshifts $10^5 < z < 10^7$ which leads to $\mu$ distortions characterized by the chemical potential, with the intensity given by Eq. \ref{eq:mu}. The chemical potential expresses energy that is absorbed or released when the number of particles in a system changes.  Here it is given in units of $kT_0$ and is therefore dimensionless.
\begin{equation} \label{eq:mu}
\Delta I_{\nu}^{\mu} = I_0\frac{x^4e^x}{(e^x-1)^2}\bigg[\frac{1}{\beta}-\frac{1}{x}\bigg]\mu,
\end{equation}
with $I_0 = \frac{2h}{c^2}(\frac{kT_0}{h})^3 = 270$~MJy/sr  for $T_0 = 2.72548$K, $\beta = 2.1923$, and $x = \frac{h \nu}{k T_0}$. 

The expected value of $\mu$ depends on various assumptions, including assumptions about possible energy injection from new physics.  Because there is no consensus about what the value \emph{should} be, we follow \cite{Abitbol2017} and \cite{Chluba2012b} in taking $\mu=2 \times 10^{-8}$ as a fiducial value (Figure \ref{fig:foreground}), in the range sometimes considered for standard $\Lambda$CDM.  This value is well below the limit detected by FIRAS of $9 \times 10^{-5}$, and in the range plausibly accessible to PIXIE.
Its zero crossing occurs at $x=\beta$, which corresponds to a frequency of 124.5~GHz.

%%%%%%%%%%%%%%%%%%%%%%%%%%%%%%%%%%%%%%%%%%%%%%%%%%%%%%%%%%%%%%%%%%%%%%%%%%%
\paragraph{Blackbody temperature distortion}
%2.7260 $\pm$ 0.0013 only wmap with FIRAS
The current estimate of the CMB blackbody temperature is  $T_0=2.72548 \pm 0.00057$K \citep{Fixsen2009}, based on the recalibration of the FIRAS data using WMAP and other measurements in the literature. PIXIE will be able to measure the blackbody temperature to a higher precision than that. This will give rise to a deviation $\Delta_T = (T_{\textrm{CMB}}-T_0)/T_0$ from the fiducial $T_0$ value, creating a spectral distortion \citep{Chluba2014a}
\begin{equation}
\Delta I_{\nu}^{\Delta_T} = I_0 \frac{x^4 e^x}{(e^x-1)^2} \Delta_T,
\end{equation}
where $I_0 = \frac{2h}{c^2}(\frac{kT_0}{h})^3 = 270$~MJy/sr  for $T_0 = 2.72548$K, and $x = \frac{h \nu}{k T_0}$.

When choosing the fiducial value for the parameter $\Delta T$, any value that is below the current measurement precision of $570\mu$K would be reasonable. For ease of comparison with \cite{Abitbol2017}, we adopt their value of $\Delta T =1.2\times 10^{-4}$K.

%%%%%%%%%%%%%%%%%%%%%%%%%%%%%%%%%%%%%%%%%%%%%%%%%%%%%%%%%%%%%%%%%%%%%%%%%%%
\paragraph{Other spectral distortions}
There are other spectral distortions that are not included in this analysis for simplicity, but that could be relevant for PIXIE, such as relativistic temperature corrections to the thermal Sunyaev-Zeldovich ($y$) distortions \citep{Sazonov1998,Abitbol2017}. 
%%%%%%%%%%%%%%%%%%%%%%%%%%%%%%%%%%%%%%%%%%%%%%%%%%%%%%%%%%%%%%%%%%%%%%%%%%%
%%%%%%%%%%%%%%%%%%%%%%%%%%%%%%%%%%%%%%%%%%%%%%%%%%%%%%%%%%%%%%%%%%%%%%%%%%%
\subsection{Foreground Modeling}\label{sec:foreground_modeling}

\paragraph{Synchrotron}
Relativistic cosmic-ray electrons are deflected by the Galactic magnetic field and emit synchrotron radiation, which is the dominant foreground at low frequency.
The \cite{Collaboration2016a} modeled this radiation as a power law with a flattening at low frequencies. \cite{Abitbol2017} allowed for a more general spectral energy distribution without using a template by fitting a power law with logarithmic curvature
\begin{equation}
\Delta I^{\textrm{s}}_{\nu} = A_{\textrm{s}} \left(\frac{\nu}{\nu_0}\right)^{\alpha_\textrm{s}}\left(1+\frac{1}{2}\omega_{\textrm{s}} \ln^2{(\frac{\nu}{\nu_0})}\right),
\end{equation}
with $\nu_0=100$~GHz. They estimated the 3 free parameters, $A_{\textrm{s}}, \alpha_\textrm{s}$, and $\omega_{\textrm{s}}$ by fitting these to the \emph{Planck} synchrotron spectrum, obtaining values of $288$~Jy/sr,$-0.82$, and $0.2$, respectively. We use these values as our fiducial values as well. 
%%%%%%%%%%%%%%%%%%%%%%%%%%%%%%%%%%%%%%%%%%%%%%%%%%%%%%%%%%%%%%%%%%%%%%%%%%%
\paragraph{Cosmic infrared background (CIB)}
Dust present in other galaxies emits thermal radiation that is redshifted on its way to us. The emission coming from all galaxies taken together is referred to as the cosmic infrared background. 
The \cite{Collaboration2014d} modeled the CIB spectral energy distribution using a modified blackbody. We take the functional form:

\begin{equation}\label{eq:CIB_MBB_no_shadow}
\Delta I^{\textrm{CIB}}_{\nu} = \alpha_{\textrm{CIB}} \left(\frac{\nu}{\nu_0}\right)^{\beta_{\textrm{CIB}}} \frac{(\nu/\nu_0)^3}{e^{h\nu/k/T_{\textrm{CIB}} }-1}
\end{equation}
with $\nu_0 = $100~GHz

$\alpha_{\textrm{CIB}}$ is related to the parameter $A_{\textrm{CIB}}$ used by \cite{Abitbol2017} via
\begin{equation}
\alpha_{\textrm{CIB}} = A_{\textrm{CIB}} \left( \frac{h \nu_0 }{kT_{\textrm{CIB}}}\right)^{\beta_{\textrm{CIB}}+3}
\end{equation}
Using their values of $\beta_{\textrm{CIB}} = 0.6$,  $T_{\textrm{CIB}} = 18.8$K, and $A_{\textrm{CIB}} =3.46\times 10^{-1}$~MJy/sr (which correpond to Planck's values as well), we obtain a fiducial value for $\alpha_{\textrm{CIB}}$ of $1.779\times 10^{-3}$~MJy/sr.

\cite{Abitbol2017} cautioned against extending the model to high frequencies, which for the purposes of this analysis we do up to $6$~THz. 
 
\cite{Nashimoto2020} highlighted the importance of modeling the absorption of the CMB's monopole component. In \S \ref{sec:CMB_shadows}, we analyze the impact of including this effect in the foreground modeling, and conclude it is necessary. Fortunately, the emissivity function (that multiplies the Planck function) is the same for both the CMB absorption and dust emission, so the required modification is straightforward:

\begin{equation}\label{eq:CIB_MBB_shadow}
\Delta I^{\textrm{CIB,s}}_{\nu} = \alpha_{\textrm{CIB}} \left(\frac{\nu}{\nu_0}\right)^{\beta_{\textrm{CIB}}} \left( \frac{(\nu/\nu_0)^3}{e^{\frac{h\nu}{kT_{\textrm{CIB}}} }-1} - \frac{(\nu/\nu_0)^3}{e^{\frac{h\nu}{kT_0} }-1} \right)
\end{equation}

% They'll get there when they get there.  :-)
% And yes, we and Naruto will be disappointed if this doesn't end up on Twitter.

%%%%%%%%%%%%%%%%%%%%%%%%%%%%%%%%%%%%%%%%%%%%%%%%%%%%%%%%%%%%%%%%%%%%%%%%%%%
\paragraph{Free-free emission}
Collisions between electrons and ions in ionized gas produce bremsstrahlung (thermal free-free) emission. Following \cite{Abitbol2017}, we use an approximation of the spectrum derived from \cite{Draine2011}. Defining $\nu_{\textrm{ff}}=\nu_{\textrm{FF}}(T_e/10^3\textrm{K})^{3/2}$, with $T_e = 7000$K and $\nu_{\textrm{FF}}=255.33$~GHz, we have
\begin{equation}
\Delta I^{\textrm{FF}}_{\nu} = A_{\textrm{FF}}(1+\ln{[1+(\frac{\nu_{\textrm{ff}}}{\nu})^{\sqrt{3}/\pi}]})
\end{equation}
In our analysis, we let only the parameter $A_{\textrm{FF}}$ vary. We use a fiducial value of $300$~Jy/sr, which \cite{Abitbol2017} obtained by fitting the model to the free-free spectrum from the \cite{Collaboration2016a}.
%%%%%%%%%%%%%%%%%%%%%%%%%%%%%%%%%%%%%%%%%%%%%%%%%%%%%%%%%%%%%%%%%%%%%%%%%%%
\paragraph{Other foregrounds not included}
In this paper we choose to focus on dust, both Galactic and extragalactic, because this already poses a significant challenge for PIXIE. In future work, we could include other foregrounds like CO emission from distant galaxies, emission from spinning dust grains, and intergalactic dust \citep{Imara2016}.

See Table \ref{table:foreground_fiducial_table} for a summary of the fiducial values for the spectral distortions and foreground parameters other than dust.

%%%%%%%%%%%%%%%%%%%%%%%%%%%%%%%%%%%%%%%%%%%%%%%%%%%%%%%%%%%%%%%%%%%%%%%%%%%
%%%%%%%%%%%%%%%%%%%%%%%%%%%%%%%%%%%%%%%%%%%%%%%%%%%%%%%%%%%%%%%%%%%%%%%%%%%
\subsection{PIXIE Mission Configuration}\label{sec:PIXIE_sensitivity}
PIXIE is expected to observe in two different configurations, aimed at polarization measurements and at intensity measurements \citep{Kogut2011}. For our analysis, we focus on the intensity measurement, and assume the mission spends 24 months in this configuration.

\cite{Kogut2020} described the current expected PIXIE sensitivity in the spectral distortion measuring mode\footnote{We obtained the table with the PIXIE sensitivity through an e-mail correspondence with Alan Kogut}. There are 416 frequency bins, each spaced 14.4~GHz apart, spanning a frequency range from 14.4~GHz to 5994.1~GHz. The covariance matrix is diagonal. For 12 months of exposure, for a 1 square degree patch in the sky, at 14.4~GHz, the table gives a value of $8.66\times10^{-24}$~W/m$^2$/sr/Hz/deg$^2=8.66\times10^{-4}$~MJy/sr/deg$^2$ (using 1~MJy = $10^6\times10^{-26}$~W/m$^2$/Hz).  There are 41253~deg$^2$ in the sky,
%\footnote{There are $4\pi$ steradians in the sky, and one steradian corresponds to $(180/\pi)^2$ degrees. Thus, there are $4\pi\times(180/\pi)^2 = 41253$ square degrees in the sky.}
of which perhaps 70\% might be available for analysis (because the foreground levels are low enough to be tractable).  They each give us an independent measurement, for $M = 0.7\times4\pi\times(180/\pi)^2= 28877$ independent measurements.  Assuming spatially uncorrelated noise, we divide the noise in each bin by $\sqrt{M}$. Next, because the starting values assume a 12 month mission and we assume a 24 month mission, we divide by $\sqrt{2}$. For 14.4~GHz, we obtain $3.60\times 10^{-6}$~MJy/sr. The obtained sensitivity-versus-frequency function is plotted in Figure \ref{fig:foreground}. It is flat at low frequency and turns up around 1~THz due to the low-pass filters in the optics. The filters are used to reduce the response to zodiacal light, which would otherwise increase the photon noise.

\begin{figure*}[t]
	\includegraphics[scale=1]{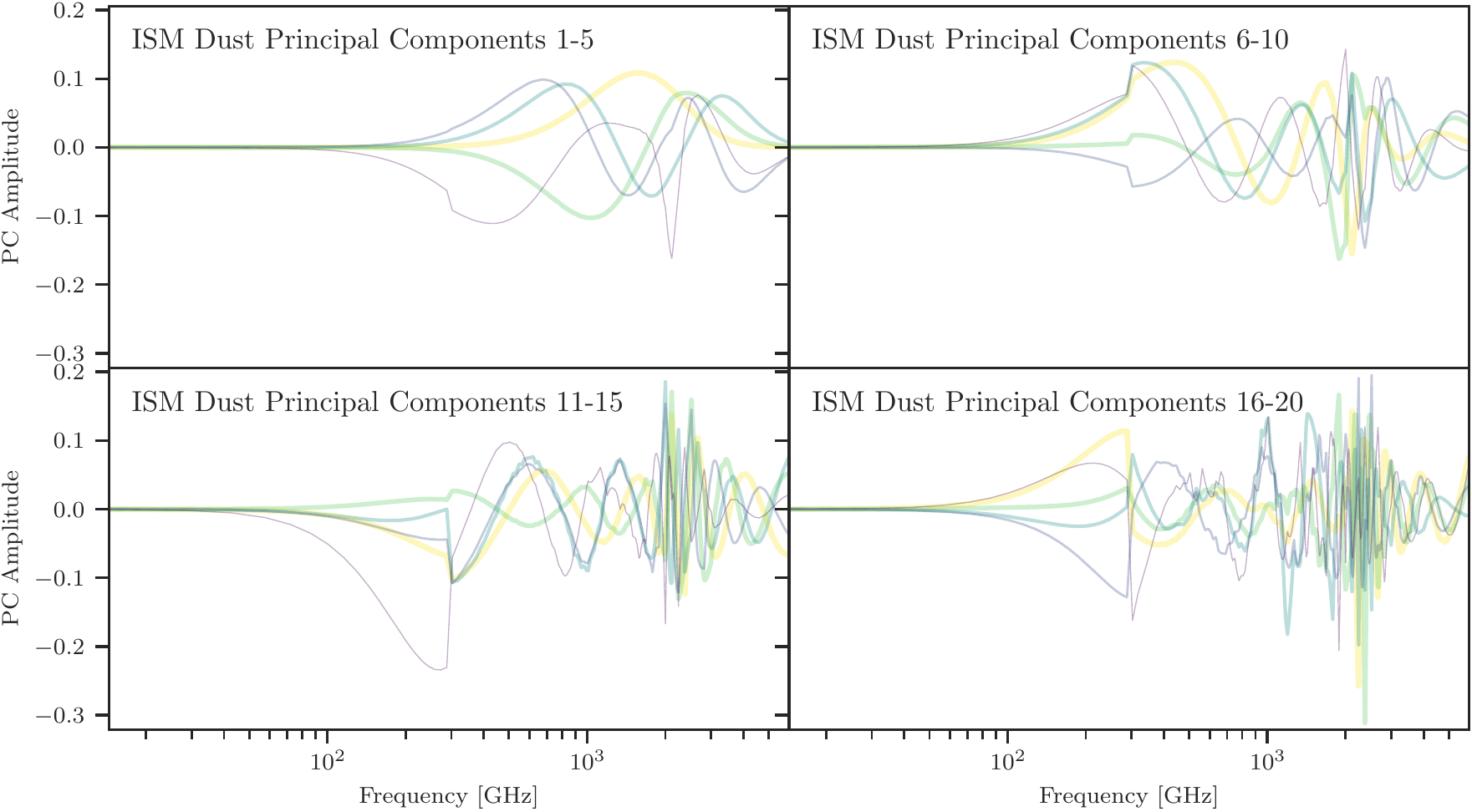}
	\caption{The first 20 principal components of ISM dust, as a function of frequency. The edges that can be seen at 300GHz most likely come from the fact that we extended the optical parameters of the dust grains beyond the 300GHz, as discussed in Appendix B of \cite{Zelko2020}. The original optical parameters come from a table with 4 significant figures \citep{Laor1993,Draine1984,Li2001}. Even if the fit got the first couple of derivatives right, there might be an edge at higher principal components because those components catch the inaccuracies in the fit.\label{fig:ISM_dust_principal_components}}
\end{figure*}

\begin{table*}[t]
\setlength{\tabcolsep}{3pt}
\centering
\footnotesize
\begin{tabular}{lrrrrrrrrrr}
\hline
PC Nr.   & 1       & 2       & 3       & 4       & 5       & 6       & 7        & 8        & 9        & 10       \\ \hline
Variance Ratio Fixed ISRF & 9.84e-01&1.59e-02&2.46e-04&4.15e-05&8.54e-07&2.11e-07&3.49e-08&1.97e-08&8.59e-09&2.09e-09 \\ \hline
Variance Ratio Variable ISRF & 9.75e-01&2.52e-02&2.59e-04&3.00e-05&3.66e-06&9.49e-07&2.62e-07&8.98e-08&2.42e-08&1.12e-08 \\ \hline
Var. Ratio Fixed ISRF No Shadow & 9.84e-01&1.59e-02&2.46e-04&4.16e-05&8.54e-07&2.12e-07&3.46e-08&1.97e-08&8.62e-09&2.05e-09 \\ \hline
Var. Ratio Variable ISRF No Shadow & 9.75e-01&2.52e-02&2.59e-04&3.00e-05&3.67e-06&9.55e-07&2.62e-07&8.97e-08&2.42e-08&1.11e-08 \\ \hline
\end{tabular}
\centering
\begin{tabular}{lrrrrrrrrrr}
\hline
PC Nr.   & 11       & 12       & 13       & 14       & 15       & 16       & 17        & 18        & 19        & 20       \\ \hline
Variance Ratio Fixed ISRF & 9.88e-10&2.07e-10&5.84e-11&4.36e-11&1.59e-11&8.68e-12&4.09e-12&1.64e-12&9.84e-13&5.84e-13 \\ \hline
Variance Ratio Variable ISRF & 1.99e-09&7.49e-10&3.27e-10&2.08e-10&4.72e-11&3.86e-11&1.54e-11&9.48e-12&6.81e-12&2.85e-12 \\ \hline
Var. Ratio Fixed ISRF No Shadow & 9.85e-10&2.06e-10&5.89e-11&4.38e-11&1.62e-11&8.65e-12&4.24e-12&1.65e-12&9.81e-13&5.84e-13 \\ \hline
Var. Ratio Variable ISRF No Shadow & 1.99e-09&7.48e-10&3.19e-10&2.05e-10&5.08e-11&3.92e-11&1.54e-11&9.50e-12&7.09e-12&2.84e-12 \\ \hline
\end{tabular}
\caption{The explained variance ratio of the first 20 principal components of the dust emissivity, for the cases of fixed and variable interstellar radiation field (ISRF). \label{table:ISM_dust_PC_variance}}
\end{table*}
\section{ISM Dust Modeling}\label{sec:dust_modeling}
The sensitivity of PIXIE to spectral distortions depends on how well the dust emission can be modeled, which in turn depends on both the correctness of the dust model and its degeneracy with $y$ and $\mu$. Our goal is to provide a better estimate of the uncertainty introduced by dust to the PIXIE spectral distortion measurement by studying the impact that varying the dust composition, size distribution, and interstellar radiation field can have on the model. 

In previous work \citep{Zelko2020}, the dust size distribution models proposed by \cite{Weingartner2001a} were used to calculate extinctions, and an MCMC was used to explore the parameter space consistent with observed extinction laws \citep{Schlafly2016}. 
For each sample point from the posterior, using precomputed values of the temperature $T$ for each radius of the grain, we integrate to calculate the specific intensity for the corresponding size distribution. Then, we calculate the principal components of the spectral energy distributions for the set of frequencies expected to be observed by PIXIE.

\subsection{Calculating the Emission Intensity from a Collection of Grains}
This section shows the procedure we use to calculate the spectral emission of a collection of dust grains given a size distribution.

The emissivity (the power radiated per unit volume
per unit frequency per unit solid angle) coming from a collection of grains is defined as:
\begin{equation}
j_{\nu}=\sum_i \int \dd a \dv{n_i}{a}C_{\textnormal{abs},i}(\nu,a)B_{\nu}(T_{\textnormal{eq}}(i,a))
\end{equation}
where $i$ is the index for carbonaceous, silicate, and polycyclic aromatic hydrocarbon grains; $C_{\textnormal{abs},i}(\nu, a)$ is the effective extinction cross section as a function of grain radius $a$; $\dv{n_i}{a}$ is the size distribution for each type of grain; and $T_{\textnormal{eq}}(i,a)$ is the equilibrium temperature of the grain of type $i$ and radius $a$.
The spectral intensity $I_{\nu}$ is defined as the emissivity integrated along the line of sight $s$:
\begin{equation}
I_{\nu} = \int j_{\nu} \dd s  = \int \left(\frac{j_{\nu}}{n_{\textnormal{H}}}\right)(s) n_{\textnormal{H}}(s) \dd s 
\end{equation}
Since $\frac{n_i}{n_{\textnormal{H}}}$ is assumed to be constant along the line of sight, $\frac{j_{\nu}}{n_{\textnormal{H}}}$ becomes constant along the line of sight as well. Using $ N_{\textnormal{H}} = \int  n_{\textnormal{H}}(s) \dd s$ , we obtain:

\begin{equation}
\begin{split}
I_{\nu} & =  \frac{j_{\nu}}{n_{\textnormal{H}}} \int  n_{\textnormal{H}}(s) \dd s = \frac{j_{\nu}}{n_{\textnormal{H}}} N_{\textnormal{H}}   = \\
& = N_{\textnormal{H}} \sum_i \int \dd a \frac{1}{n_{\textnormal{H}}}\dv{n_i}{ a} C_{\textnormal{abs},i}(\nu,a)B_{\nu}(T_{\textnormal{eq}}(i,a))
\end{split}
\end{equation}

\cite{Nashimoto2020} demonstrated the importance of modeling the absorption of the CMB monopole (see \S \ref{sec:CMB_shadows}). In this work we take this effect into account, and consider the net dust emission (subtracting the absorbed CMB):

\begin{equation}
\begin{split}
I_{\nu} & =  \frac{j_{\nu}}{n_{\textnormal{H}}} \int  n_{\textnormal{H}}(s) \dd s = \frac{j_{\nu}}{n_{\textnormal{H}}} N_{\textnormal{H}}   = \\
& = N_{\textnormal{H}} \sum_i \int \dd a \frac{1}{n_{\textnormal{H}}}\dv{n_i}{ a} C_{\textnormal{abs},i}(\nu,a)\left(B_{\nu}(T_{\textnormal{eq}}(a)) - B_{\nu}(T_0)\right)
\end{split}
\end{equation}
where $T_0$ is the CMB temperature.

%

%%%%%%%%%%%%%%%%%%%%%%%%%%%%%%%%%%%%%%%%%%%%%%%%%%%%%%%%%%%%%%%%%%%
\subsection{Principal Component Analysis}

For each sample point from \cite{Zelko2020} with its  corresponding size distribution,  using the precomputed values of the temperature $T$ for each radius and type of grain, we integrate to calculate the specific intensity
at each of the 416 PIXIE frequencies (Figure \ref{fig:foreground}). 
Thinking of each such spectrum as a point in a 416-dimensional space, we can use PCA \citep{Shlens2014} to find a low-dimensional subspace that contains most of the variance.  In other words, the ensemble of models from \cite{Zelko2020} were generated from an 11-parameter size distribution and a range of ISRF heating, but they span a relatively low-dimensional ``dust space'' of possible emission spectra.  To the extent that the $\mu$ and $y$ spectral distortions are orthogonal to this dust space, they can be measured by PIXIE.  

When computing the principal components, we first scale the dust emissivity by the PIXIE sensitivity, as described in \S \ref{sec:PIXIE_sensitivity}. Then, we subtract the average.

To reconstruct the spectrum from the principal components, one uses the Eq.:
\begin{equation}
\Delta I_{\nu}^{\textrm{dust}} = \left(\sum_{i} W_i PC_i +\textrm{mean}\right)\times \sigma_{\nu}^{\textrm{PIXIE}}
\end{equation}

We perform two different analyses: one in which the interstellar radiation field is fixed, and another where we multiply the interstellar radiation field by a factor $\chi_{\textrm{ISRF}}$ varying between 0.5 and 2, which generates different equilibrium temperatures for the grains, and thus a different SED.

The 20 principal components we obtain can be seen in Figure \ref{fig:ISM_dust_principal_components}. Table \ref{table:ISM_dust_PC_variance}  shows the variance explained by each principal component. We see that in both cases, the first component explains most of the variance, but the relative importance of it increases when we let the radiation field vary.

\begin{table}[t]
\centering
\begin{tabular}{|r|r|r|r|r|r|}
\toprule
PC Parameter & WC0 & WC1 & WC2 & WC3  \\
\hline
Fiducial Value&8.3e+05&-2.4e+05&-3.6e+03&-6.7e+03 \\
\hline
\end{tabular}%
\caption{Fiducial values for the PCA coefficients used in the MCMC. The coefficients are obtained by decomposing a reference SED (shown in Fig. \ref{fig:foreground}) from the ISM dust posteriors from \cite{Zelko2020} unto the PCs. \label{table:PCA_initial}}
\end{table}

\begin{figure*}[t!]
    \hspace{-7mm}
	\includegraphics[scale=0.25]{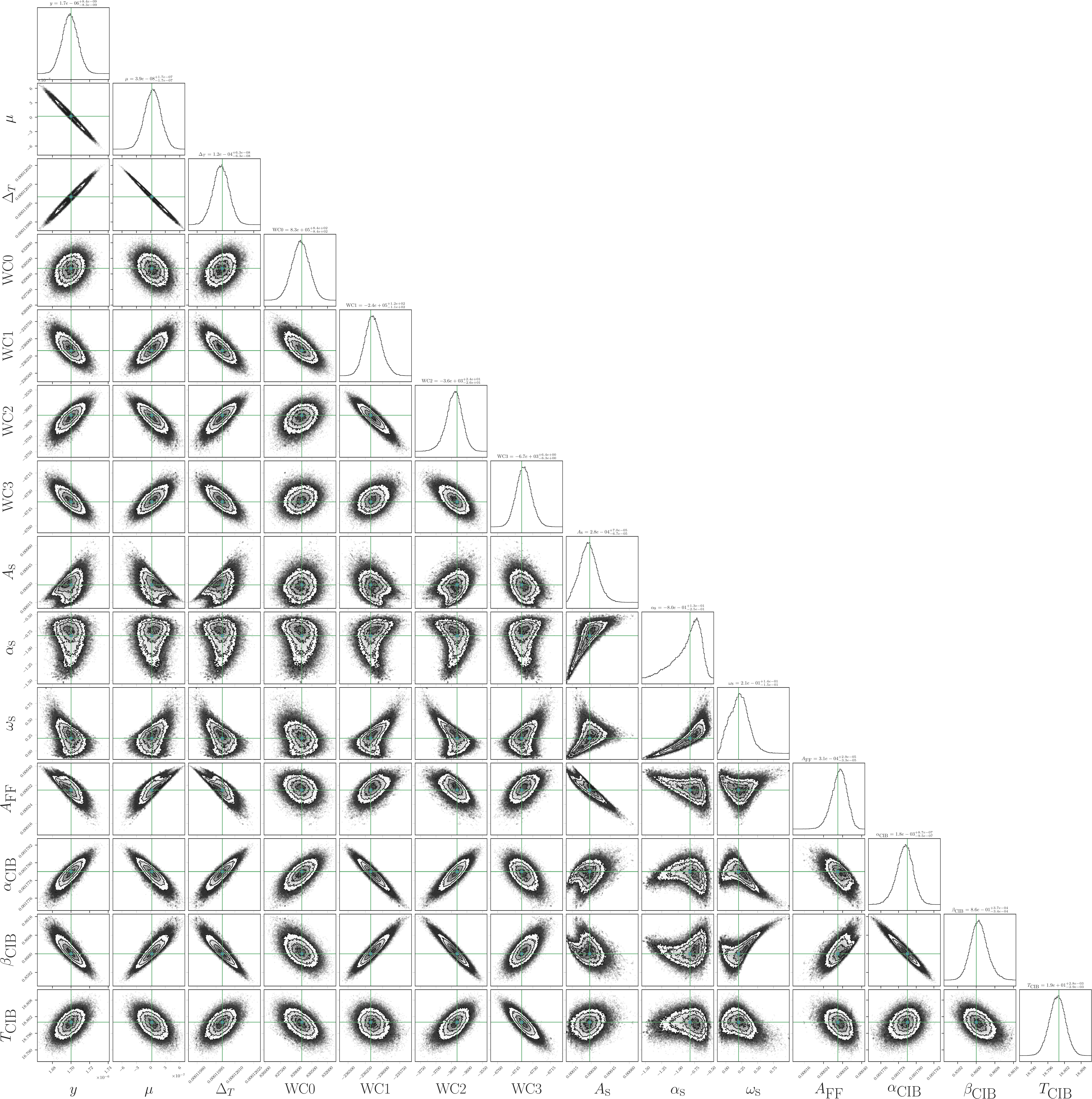}
	\caption{Results of the MCMC exploring the posterior space for the parameters describing the spectral distortions in the presence of ISM dust modeled by 4 principal components, synchrotron radiation, free-free emission, and the CIB. We see that with the exception of synchrotron, the posteriors are Gaussian. For synchrotron radiation, the posterior is Gaussian at 1-$\sigma$ level. \label{fig:MCMC_ISM_dust_PC}}
\end{figure*}

\begin{table*}[t]
\setlength{\tabcolsep}{1.5pt}
\centering
\scriptsize
\begin{tabular}{rrrrrrrrrrrrrrr}
\hline
Param Name & $y$&$\mu$&$\Delta_T$&WC0&WC1&WC2&WC3&$A_{\textrm{s}}$&$\alpha_{\textrm{s}}$&$\omega_{\textrm{s}}$&$A_{\textrm{FF}}$&$\alpha_{\textrm{CIB}}$&$\beta_{\textrm{CIB}}$&$T_{\textrm{CIB}}$ \\ \hline
Units & -&-&[K]&-&-&-&-&[MJy/sr]&-&-&[MJy/sr]&[MJy/sr]&-&[K] \\ \hline
Fiducial Val. & 1.70e-06&2.00e-08&1.20e-04&8.30e+05&-2.36e+05&-3.63e+03&-6.74e+03&2.88e-04&-8.20e-01&2.00e-01&3.00e-04&1.78e-03&8.60e-01&1.88e+01 \\ \hline
Post. Median & 1.70e-06&3.95e-08&1.20e-04&8.29e+05&-2.36e+05&-3.64e+03&-6.74e+03&2.77e-04&-8.03e-01&2.11e-01&3.07e-04&1.78e-03&8.60e-01&1.88e+01 \\ \hline
$\sigma_{+}$ & 8.41e-09&1.66e-07&6.37e-08&8.41e+02&1.22e+02&2.41e+01&6.48e+00&7.01e-05&1.34e-01&1.45e-01&2.97e-05&8.77e-07&3.74e-04&2.82e-03 \\ \hline
$\sigma_{-}$ & 8.33e-09&1.69e-07&6.37e-08&8.43e+02&1.14e+02&2.62e+01&6.29e+00&6.77e-05&2.48e-01&1.54e-01&3.35e-05&9.57e-07&3.46e-04&2.91e-03 \\ \hline
\end{tabular}
\caption{The results of the MCMC for the configuration checking the posterior is Gaussian for all foregrounds and 4 PC ISM dust. \label{table:foregrounds_MCMC}$ \sigma_{+}$ represents the difference between the 84th and 50th quantile of the posterior, and $\sigma{-}$ the one between the 50th and 16th.}
\end{table*}

\begin{table*}[ht]
\centering
\setlength{\tabcolsep}{1.1pt}
\begin{tabular*}{\linewidth}{ @{\extracolsep{\fill}}r *{10}{r} @{} }
\toprule
 & &\multicolumn{9}{c@{}}{Number of PCs used} \\
\cmidrule(l){3-11}
 & MBB & 2PC & 3PC & 4PC & 5PC & 6PC & 7PC & 8PC & 9PC & 10PC   \\
\midrule
\multicolumn{6}{@{}l@{}}{\textit{Fixed radiation field}} \\
$\sigma{y}$ &4.81e-09&5.85e-09 & 6.02e-09 & 6.02e-09 & 9.31e-09 & 9.33e-09 & 9.35e-09 & 1.02e-08 & 1.10e-08 & 1.23e-08 \\
$\sigma{\mu}$ &8.28e-08&1.11e-07 & 1.15e-07 & 1.16e-07 & 1.79e-07 & 1.81e-07 & 1.82e-07 & 1.96e-07 & 2.42e-07 & 2.72e-07 \\
\midrule
\multicolumn{6}{@{}l@{}}{\textit{Variable radiation field}} \\
$\sigma{y}$ &4.81e-09&4.41e-09 & 5.58e-09 & 8.89e-09 & 9.62e-09 & 1.18e-08 & 1.20e-08 & 1.30e-08 & 1.67e-08 & 1.78e-08 \\
$\sigma{\mu}$ &8.28e-08&8.36e-08 & 1.14e-07 & 1.77e-07 & 2.05e-07 & 2.63e-07 & 2.63e-07 & 2.83e-07 & 3.96e-07 & 4.09e-07 \\
\bottomrule
\end{tabular*}
\begin{tabular*}{\linewidth}{ @{\extracolsep{\fill}} r *{10}{r} @{} }
\toprule
 & &\multicolumn{9}{c@{}}{Number of PCs used} \\
\cmidrule(l){3-11}
 & 11PC & 12PC & 13PC & 14PC & 15PC & 16PC & 17PC & 18PC & 19PC & 20PC   \\
\midrule
\multicolumn{6}{@{}l@{}}{\textit{Fixed radiation field}} \\
$\sigma{y}$ &1.41e-08 & 1.46e-08 & 1.49e-08 & 2.15e-08 & 2.28e-08 & 2.44e-08 & 2.68e-08 & 2.77e-08 & 2.82e-08 & 2.86e-08 \\
$\sigma{\mu}$ &3.15e-07 & 3.32e-07 & 3.38e-07 & 5.36e-07 & 5.87e-07 & 6.14e-07 & 6.77e-07 & 7.20e-07 & 7.29e-07 & 7.44e-07 \\
\midrule
\multicolumn{6}{@{}l@{}}{\textit{Variable radiation field}} \\
$\sigma{y}$ &1.82e-08 & 1.86e-08 & 2.02e-08 & 2.48e-08 & 2.98e-08 & 3.54e-08 & 3.57e-08 & 3.57e-08 & 3.82e-08 & 3.91e-08 \\
$\sigma{\mu}$ &4.14e-07 & 4.30e-07 & 4.74e-07 & 5.94e-07 & 8.08e-07 & 9.40e-07 & 9.44e-07 & 9.45e-07 & 9.95e-07 & 1.02e-06 \\
\bottomrule
\end{tabular*}
\caption{Fisher Information Matrix analysis results. Standard deviations for the $y$ and $\mu$ parameters, given inclusion of various numbers of principal components. The case for constant and variable radiation field is shown. All other foregrounds are included in the analysis.  \label{table:results_table_1}}
\end{table*}

\begin{table*}[ht]
\setlength{\tabcolsep}{2.3pt}
\centering
\footnotesize
\begin{tabular}{rrrrrrrrrrrrrrr}
\hline
Param. &  $y \times 10^9$ & $\mu\times 10^9$& $\Delta_T$ &WC0&WC1&WC2&WC3&$ A_{\textnormal{s}}$ & $\alpha_{\textnormal{s}}$&$\omega_\textnormal{s}$ & $\alpha_{\textnormal{CIB}}$ & $\beta_{\textnormal{CIB}} $ & $T_{\textnormal{CIB}}$ & $A_{\textnormal{FF}}$  \\
\hline
Unit &  - & - & [nK] &  - &  - &  - &  - &[MJy/sr] &-& - & [MJy/sr] & - & [K] & [MJy/sr]  \\
\hline
Fid. & 1700&20&1.2e+05&8.3e+05&-2.4e+05&-3.6e+03&-6.7e+03&2.9e-04&-8.2e-01&2.0e-01&1.8e-03&8.6e-01&1.9e+01&3.0e-04 \\ \hline
$\sigma$ &0.9&9.8&1.7&1.0e+00&1.0e+00&1.0e+00&1.0e+00&-&-&-&-&-&-&- \\ \hline
$\sigma$ &1.5&23.9&5.4&-&-&-&-&3.0e-06&2.0e-03&6.5e-03&-&-&-&- \\ \hline
$\sigma$ &1.1&10.1&1.8&-&-&-&-&-&-&-&1.7e-08&6.6e-06&3.9e-05&- \\ \hline
$\sigma$ &1.0&11.5&2.3&-&-&-&-&-&-&-&-&-&-& 2.5e-07 \\ \hline
$\sigma$ &2.0&27.7&7.2&4.7e+00&2.1e+00&2.2e+00&1.5e+00&4.0e-06&9.9e-03&9.9e-03&-&-&-&- \\ \hline
$\sigma$ &1.5&10.8&2.6&7.6e+02&6.3e+01&1.1e+01&4.1e+00&-&-&-&2.9e-07&7.6e-05&2.4e-03&- \\ \hline
$\sigma$ &1.7&20.6&5.3&3.0e+00&1.5e+00&1.5e+00&1.2e+00&-&-&-&-&-&-& 8.3e-07 \\ \hline
$\sigma$ &2.2&33.9&11.1&-&-&-&-&7.3e-06&3.0e-02&2.7e-02&9.5e-08&2.9e-05&1.5e-04&- \\ \hline
$\sigma$ &1.7&34.1&8.6&-&-&-&-&3.8e-05&9.2e-02&5.6e-02&-&-&-& 1.3e-05 \\ \hline
$\sigma$ &1.5&20.7&5.3&-&-&-&-&-&-&-&2.5e-08&8.6e-06&4.9e-05& 9.2e-07 \\ \hline
$\sigma$ &7.4&140.5&58.8&8.8e+02&1.4e+02&2.7e+01&6.2e+00&4.5e-05&2.1e-01&1.9e-01&1.0e-06&4.1e-04&2.9e-03&- \\ \hline
$\sigma$ &3.1&62.9&17.3&5.3e+00&2.3e+00&2.4e+00&1.5e+00&8.2e-05&2.0e-01&1.2e-01&-&-&-& 2.9e-05 \\ \hline
$\sigma$ &2.0&22.0&6.0&7.6e+02&6.3e+01&1.1e+01&4.1e+00&-&-&-&2.9e-07&7.6e-05&2.4e-03& 9.5e-07 \\ \hline
$\sigma$ &2.3&44.6&12.9&-&-&-&-&6.7e-05&1.7e-01&1.1e-01&1.0e-07&3.1e-05&1.6e-04& 2.2e-05 \\ \hline
$\sigma$ &8.9&177.0&69.4&8.8e+02&1.4e+02&2.8e+01&6.6e+00&8.6e-05&2.6e-01&2.0e-01&1.1e-06&4.3e-04&3.0e-03& 3.1e-05 \\ \hline
\end{tabular}
\caption{Fisher information matrix estimates of uncertainty in $y$ and $\mu$ (full-sky average), allowing various combinations of foreground parameters to float (\S \ref{sec:fisherresults}).  In each row, the $1\sigma$ marginalized uncertainty is given for floating parameters, and fixed parameters are indicated by ``-''.  ISM dust is represented by 4 principal components, with emission spectra drawn from the \cite{Zelko2020} posterior (variable radiation field case).  Because the spectra are scaled by the PIXIE sensitivity in the PCA, the uncertainty of the PC coefficients (WCn) is 1.0 for the dust-only case, and larger for the other cases.  As more foreground parameters are allowed to float, the uncertainty in $y$ and $\mu$ grows.  When all foregrounds are included (bottom row), $\sigma_\mu$ is an order of magnitude larger than a case with fixed synchrotron parameters (3rd from bottom).  This suggests that independent constraints on the synchrotron spectrum would  substantially enhance the statistical power of PIXIE.  \label{table:combos_table}}
\end{table*}
\section{ISM Dust Forecasting Methods and Results} \label{sec:ISM_dust_forecasting_methods}

Our goal is to use the dust SED spectral space encoded in the principal components (\S \ref{sec:dust_modeling}) to assess the detectability of the spectral distortions of the CMB.  The Fisher information matrix provides a quick way to estimate uncertainties, assuming the likelihood is Gaussian near its peak. To confirm the validity of this assumption, an MCMC analysis is performed to visualize and confirm the actual posterior space for a sample case. These two methods are applied to the case where we assume our fit models are the models that generated the reference data. Thus, the analysis sheds light on the impact the ISM dust PCs have on the error bars placed on the $y$ and $\mu$ distortions. 
In the last subsection, we explore the case where the models used in a fit for the ISM dust are not the models that generated the reference data, and quantify the model discrepancy error. 

%%%%%%%%%%%%%%%%%%%%%%%%%%%%%%%%%%%%%%%%%%%%%%%%%%%%%%%%%%%%%%%%%%%
\subsection{MCMC for Foreground and Spectral Distortion Modeling Analysis} \label{sec:MCMC_foreground}
%\ref{sec:MCMC_foreground}
A reference intensity is created from all the spectral distortions and foregrounds: 
\begin{equation}
\Delta I_{\nu} = \Delta I_{\nu}^{\textrm{dust}} + \Delta I_{\nu}^y + \Delta I_{\nu}^{\mu} +\Delta I_{\nu}^{\Delta_T}+ \Delta I_{\nu}^{\textrm{sync}}+ \Delta I_{\nu}^{\textrm{CIB}}+ \Delta I_{\nu}^{\textrm{FF}}
\end{equation}
To generate it, for the spectral distortions and foregrounds other than dust we use the fiducial values in Table \ref{table:foreground_fiducial_table}. For the coefficients of the principal components of the dust, see Table \ref{table:PCA_initial}. 

The MCMC likelihood is defined as  $\ln{\mathcal{L}} = -0.5 \times \Delta \chi^2$, with $\Delta \chi^2 = \sum_{\nu}((\Delta I_{\nu}^{\textrm{total}}-\Delta I_{\nu}^{\textrm{reference}})/\sigma_{\nu}^{\textrm{PIXIE}})^2$, where the $\Delta I_{\nu}^{\textrm{reference}}$ is the total intensity calculated with the fiducial parameters, as described above, and $\Delta I_{\nu}^{\textrm{total}}$ is the intensity obtained as we let the MCMC explore different parameter ranges.

No priors are imposed on the parameters of the MCMC, beyond just physical boundaries. 

The following information applies to all the MCMC runs in this paper, including the ones in \S \ref{sec:CMB_shadows}: the MCMC uses the \texttt{ptemcee} \footnote{The code can be found at the Python repository at \url{https://pypi.org/project/ptemcee/} or at the Will Vousden Github repository at \url{https://github.com/willvousden/ptemcee}} \cite{Vousden2016} package, which uses parallel tempering. This allows for a more efficient exploration of the parameter space than something like the Metropolis-Hastings algorithm.  We used 5 temperatures and 300 walkers, running for 50000 steps. The runs were thinned by a factor of 50 to reduce autocorrelation. To discard the burn-in phase of the chain, only the last 50\% of the steps were retained.

\subsection{Foreground MCMC Results}

The posteriors appear to be reasonably Gaussian for all parameters except for the synchrotron radiation (Fig. \ref{fig:MCMC_ISM_dust_PC}, Table \ref{table:foregrounds_MCMC}). This is in agreement with \cite{Abitbol2017}, who obtained non-Gaussian distributions for the case where they did not impose priors on the synchrotron background, and only obtained Gaussian posteriors once the synchrotron priors were in place. Still, at the 1$\sigma$ level, the synchrotron posteriors are also Gaussian, so this validates our use of the Fisher information matrix.

\subsection{Fisher Information Matrix Analysis}\label{sec:fisher}

Assuming a Gaussian likelihood, one can use the Fisher matrix formalism to calculate the parameter uncertainties. Our likelihood takes the form
\begin{equation}
L = \frac{1}{(2\pi)^{n/2}|\det{C}|^{1/2}}\exp\Big[-\frac{1}{2}\sum_{ij} (p_i - p_{0,i})C_{ij}^{-1}(p_j - p_{0,j}) \Big],
\end{equation}
where the 0 subscript indicates the true parameter values. 
Near the peak of the likelihood, we can make a Taylor series expansion:
\begin{equation}
\log{L} = \log{L(\mathbf{p_0})} + \frac{1}{2}\sum_{ij}(p_i-p_{0,i})\frac{\partial^2\log{L}}{\partial p_i\partial p_j}\bigg\rvert_{\mathbf{p_0}}(p_j -p_{0,j}) + \dots
\end{equation}
Taking the average over the random independent variable, the second factor becomes:
\begin{equation}
F_{ij} = -\Big\langle\frac{\partial^2 \log L}{\partial p_i \partial p_j} \Big\rangle
\end{equation}

The Fisher matrix is related to the PIXIE noise covariance, $C_{ab}$, via
\begin{equation}
F_{ij} = \sum_{a,b} \frac{\partial (\Delta I_{\nu})_a}{\partial p_i} C_{ab}^{-1}\frac{\partial \Delta (I_{\nu})_b}{\partial p_j}
\end{equation}
which is called the Fisher information matrix \citep[see for example][]{Verde2010}. $a,b$ are indexing over frequency, and $p_i, p_j$ over the different parameters of the model, which in our case are the free parameters for the spectral distortions and the foregrounds. Inverting the $F_{ij}$ matrix and taking the square root of the diagonal gives us the standard deviations expected for each parameter. 

To calculate $F_{ij}$, we calculate all the derivatives of the spectral energy distributions with respect to the free parameters. In our case, they take a simple analytic form, as described in the Appendix  \ref{sec:derivative_calculation}.

%%%%%%%%%%%%%%%%%%%%%%%%%%%%%%%%%%%%%%%%%%%%%%%%%%%
\subsection{Fisher Information Matrix Results}\label{sec:fisherresults}

\begin{table}[t]
	\centering
	\begin{tabular}{cccc}
		\toprule
		\multicolumn{4}{c}{4 ISM dust PC Components+ all other foregrounds} \\
		\toprule
		\multicolumn{2}{l|}{Fisher Analysis Results} & \multicolumn{2}{l}{MCMC Analysis Results} \\
		\multicolumn{1}{l}{	$y$} & \multicolumn{1}{c|}{8.89e-09 }&\multicolumn{1}{l}{$y$}& \multicolumn{1}{c}{8.37e-09 } \\
		\multicolumn{1}{l}{	$\mu$}& \multicolumn{1}{c|}{1.77e-07} & \multicolumn{1}{l}{$\mu$} & \multicolumn{1}{c}{1.68e-07} \\
		\bottomrule
	\end{tabular}
	\caption{Comparison between Fisher Information and MCMC results for the ISM dust 4 PC modeling, in the presence of other foregrounds. \label{table:MCMC_vs_Fisher}}
\end{table}

As the MCMC approach showed us that we can safely use the Fisher information matrix approach, we perform it using different numbers of principal components, to see what effect complicating the dust model has on the detection sensitivity. The results are displayed in Table  \ref{table:results_table_1}. One of the principal questions that this work is aiming to answer is how much does the sensitivity to $\mu$ and $y$ degrades as a function of the complexity of the dust model. We observe that  most of the PIXIE noise-weighted variance in spectral space is contained in the first two principal components, and adding up to 20 principal components increases the standard deviation for the $y$ and $\mu$ distortion parameters by less than one order of magnitude.

We explore all the combinations of foregrounds and see what impact they have on the standard deviation of $y$ and $\mu$. The results are summarized in Table \ref{table:combos_table}. Adding all foregrounds one by one produces an increase in the expected standard deviation of the $y$ and $\mu$ parameters, with the highest $\sigma$ expected when all foreground parameters need to be constrained at once. When the foregrounds are modeled only one at a time (leaving the others to be constrained by dedicated experiments), it can be seen that dust produces the lowest impact on the detectability, followed by the cosmic infrared background, the free-free emission, and finally the synchrotron radiation which produces the most impact. In particular, the $\mu$ signal could only be detected if it is stronger than the fiducial value of $2\times10^{-8}$. However, if the synchrotron background is first constrained by a dedicated experiment, we can have a good possibility of detecting it in the range sometimes considered for standard $\Lambda$CDM of $2\times10^{-8}$, even in the presence of our dust model.

In order to validate our methods, we compare the results from the MCMC and the Fisher information matrix. We perform both analyses using the same conditions, modeling the foregrounds for dust, synchrotron radiation, CIB  and free-free emission. The results are listed in Tables \ref{table:foregrounds_MCMC} and \ref{table:MCMC_vs_Fisher}, where it can be seen that the two methods are in agreement.

\begin{figure}[ht!]
	\includegraphics[scale=1]{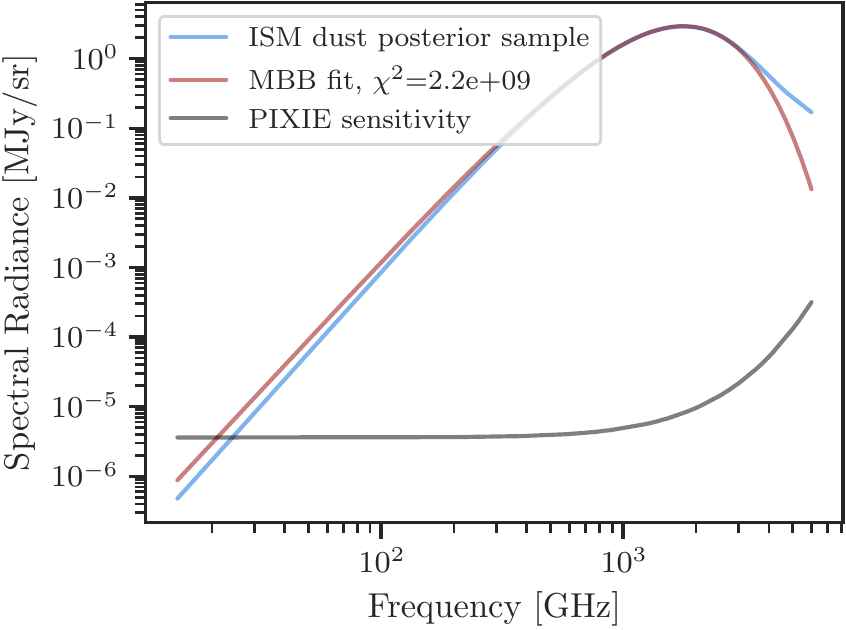}
	\caption{Example of an MBB fit to the ISM dust SED constructed from a single sample from the posterior obtained in \cite{Zelko2020}. When fit using the sensitivity of PIXIE, the $\chi^2$ becomes too big. \label{fig:MBB_to_posterior_fit}}
\end{figure}

\begin{figure*}[t!]
	\centering
	\begin{tabular}{cc}
		\hspace{-5.00mm}
		\includegraphics[width=0.48\textwidth]{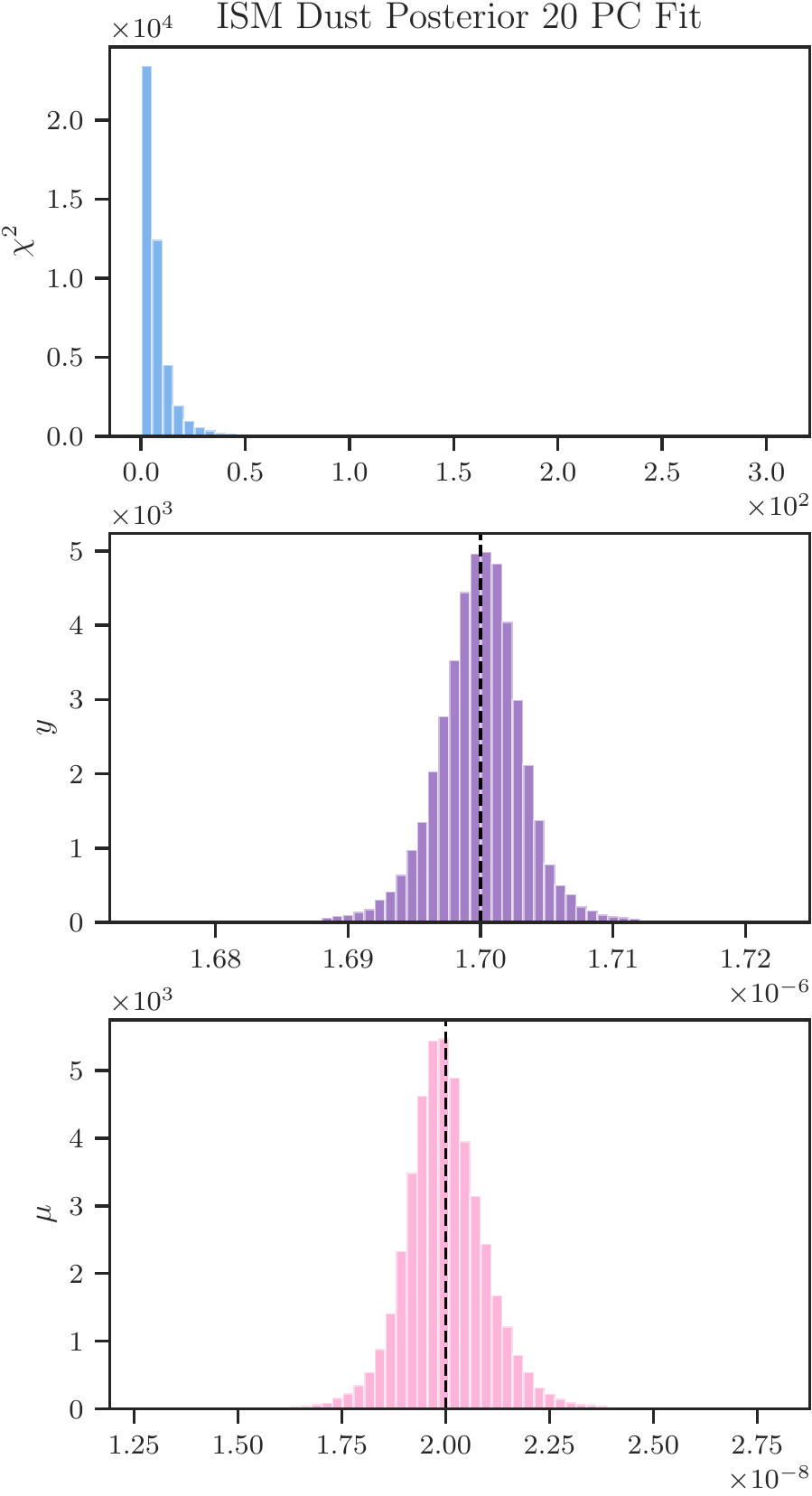} &
		\hspace{-2.00mm}	
		\includegraphics[width=0.48\textwidth]{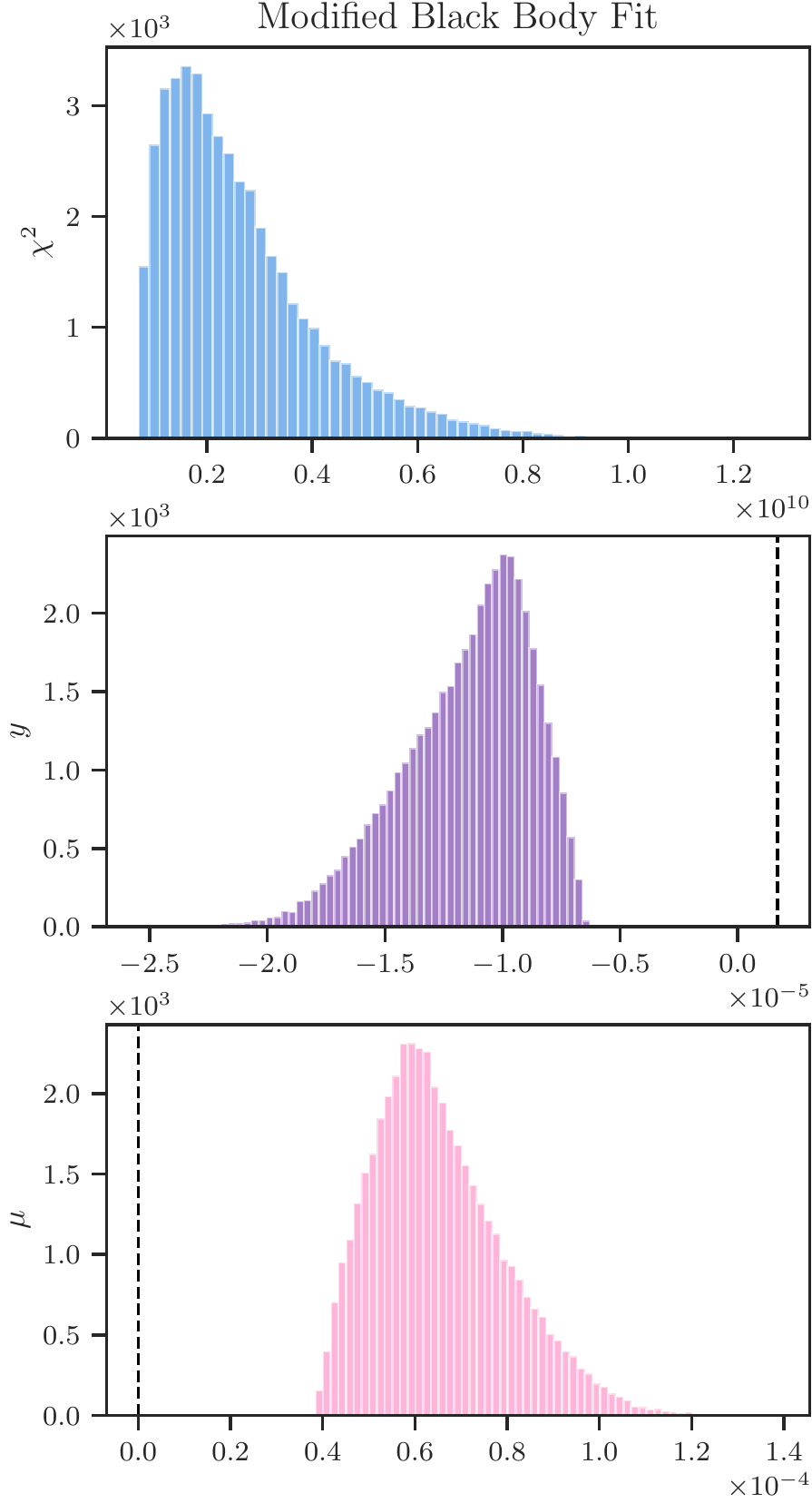} 
		
	\end{tabular}
	
	\caption{We perform fits for each of the 45000 SEDs obtained from the posterior of \cite{Zelko2020}. The panels on the left column show fits performed using 20 ISM dust PCs. One can see that they reproduce the data well, and there is no impact on the $y$ and $\mu$ parameters. The right panels show fits performed using a MBB. Since the model does not fit the ISM dust data, it results in biases in the $y$ and $\mu$ parameters.	
		\label{fig:PC_fits_vs_MBB_fits}}
	\end{figure*}

\subsection{Model Discrepancy Error}
The results presented above using the Fisher information matrix and MCMC are useful in determining the impact ISM dust PCs have on the $\sigma_{y}$ and $\sigma_{\mu}$. However, by the nature of the fits, they were testing the situation in which the correct model is reproducing the data. So the question arises, what happens if the wrong model is used to model the ISM dust? An example of such a case is the following.  We take as ground truth the ISM dust SED obtained from the size distributions of the dust grains from a sample of the posteriors in \cite{Zelko2020}. As the model for the fit, an MBB is used. The goal is to see what biases are introduced to the $y$ and $\mu$ distortions.

To fit the ISM SED, an MBB with the CMB shadow contribution is used:
\begin{equation}\label{eq:MBB_ISM_dust_shadow}
\Delta I_{\nu}^\textrm{dust MBB,s} = \tau_{353} \big ( \frac{\nu}{353\text{ GHz}} \big )^{\beta} \left(B_{\nu}(T_{\textrm{D}})-B_{\nu}(T_0) \right) 
\end{equation}
where $T_{\textrm{D}}$ is the dust temperature, and $T_0$ is the CMB temperature.

We show the results of performing fits for each of the 45000 SEDs obtained from the posterior of \cite{Zelko2020} in Fig. \ref{fig:PC_fits_vs_MBB_fits}. Fits with 20PCs from the ISM dust posterior were also performed for comparison. It can be seen that the difference in fits is staggering. Of course, the PCs were by definition obtained on this SED space, so including a sufficient number of them should produce good fits. The case of the MBB introduces biases to both $y$ and $\mu$ distortions that make detecting their real values impossible. This is due to the fact that at the PIXIE sensitivity levels, an MBB is a very poor fit to the ISM dust SED created from the dust grain size distributions, as shown in Fig. \ref{fig:MBB_to_posterior_fit}. No other foregrounds were included in the analysis.

In summary, an overly simplistic dust model like an MBB will bias spectral distortion measurements at a level orders of magnitude greater than the expected sensitivity of PIXIE. This could potentially be worked around by using  spatial information when considering the variation of SED parameters using ILC and moment expansion techniques \citep{Chluba2017, Remazeilles2020, Rotti2021}.

\begin{figure}[t]
	\includegraphics[scale=1]{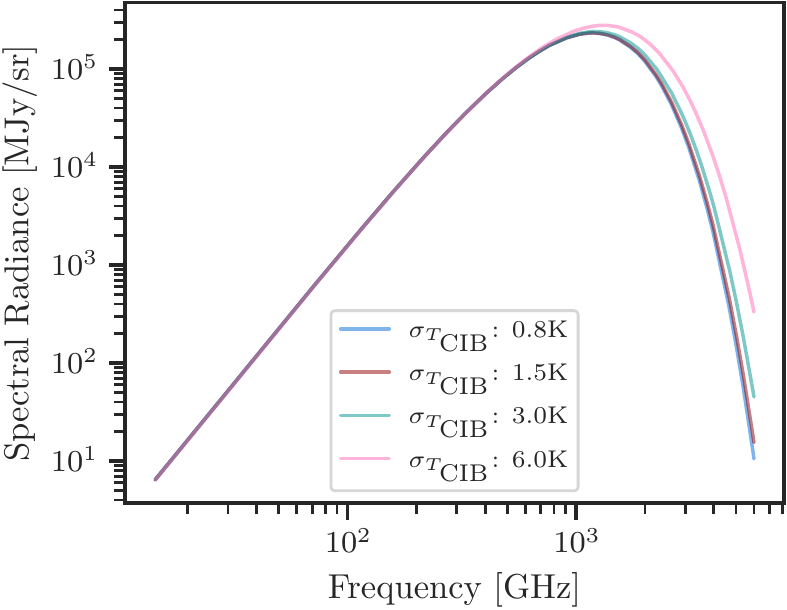}
	\caption{Examples of CIB SEDs created from superposition of MBB with  $T_{\textrm{CIB}}$ drawn from a Gaussian distribution at different $\sigma_{T_{\textrm{CIB}}}$\label{fig:CIB_broadening_example}}
\end{figure}

\begin{figure*}[t]
	\includegraphics[scale=1]{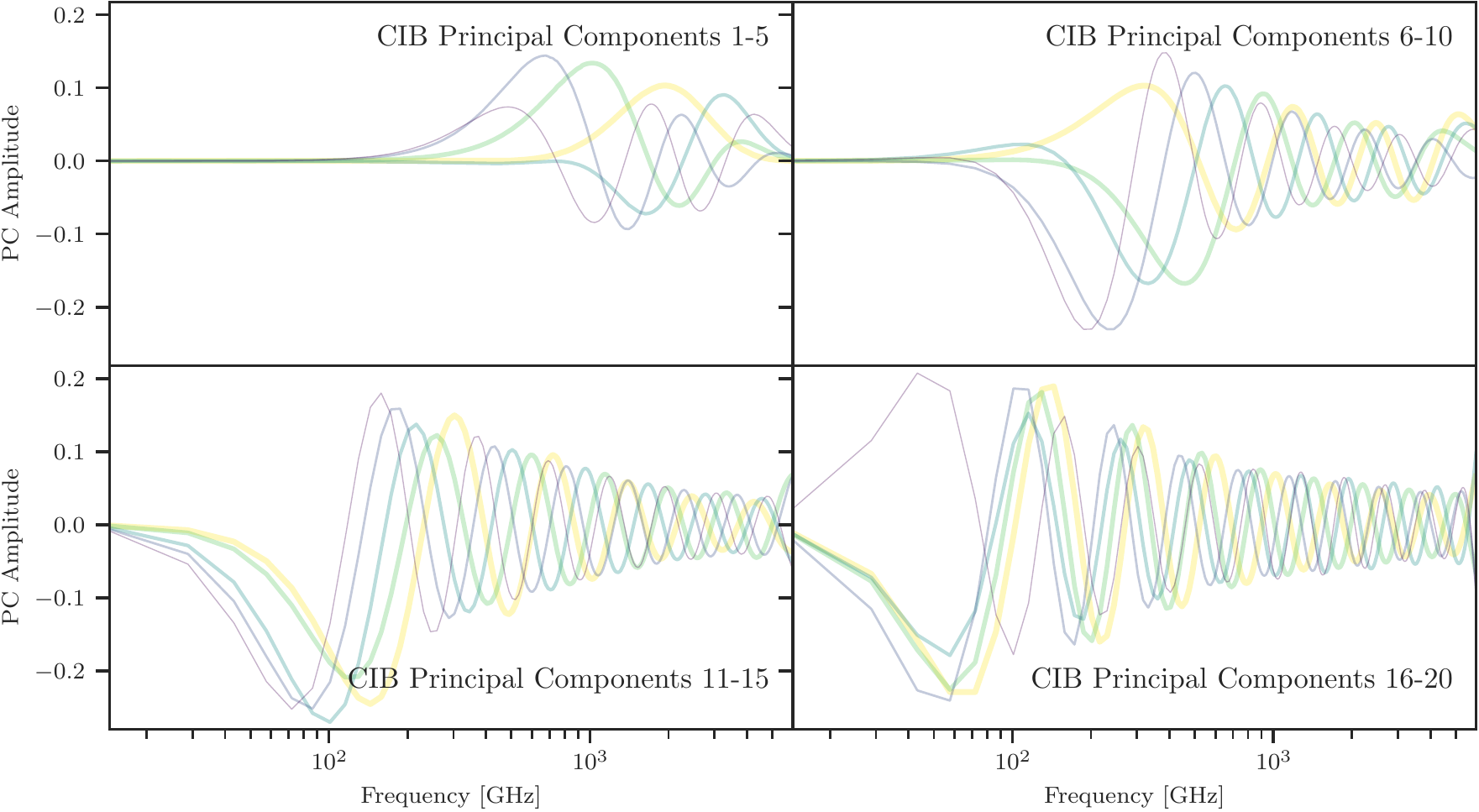}
	\caption{The first 20 principal components of CIB, as a function of frequency.\label{fig:CIB_principal_components}}
\end{figure*}
\begin{table*}[t]
\setlength{\tabcolsep}{3pt}
\centering
\footnotesize
\begin{tabular}{lrrrrrrrrrr}
\hline
PC Nr.   & 1       & 2       & 3       & 4       & 5       & 6       & 7        & 8        & 9        & 10       \\ \hline
PC Variance& 9.99e-01&6.16e-04&5.01e-05&1.51e-07&4.02e-09&1.59e-11&6.34e-12&1.56e-13&4.07e-14&4.95e-15 \\ \hline
\end{tabular}
\centering
\begin{tabular}{lrrrrrrrrrr}
\hline
PC Nr.   & 11       & 12       & 13       & 14       & 15       & 16       & 17        & 18        & 19        & 20       \\ \hline
PC Variance & 1.40e-15&1.50e-16&2.86e-17&3.11e-18&3.66e-19&4.39e-20&4.08e-21&4.63e-22&4.56e-23&4.41e-24 \\ \hline
\end{tabular}
\caption{The explained variance ratio of the first 20 principal components of the broadened CIB dataset. \label{table:CIB_PC_variance}}
\end{table*}

\begin{figure*}[ht!]
	\includegraphics[scale=0.94]{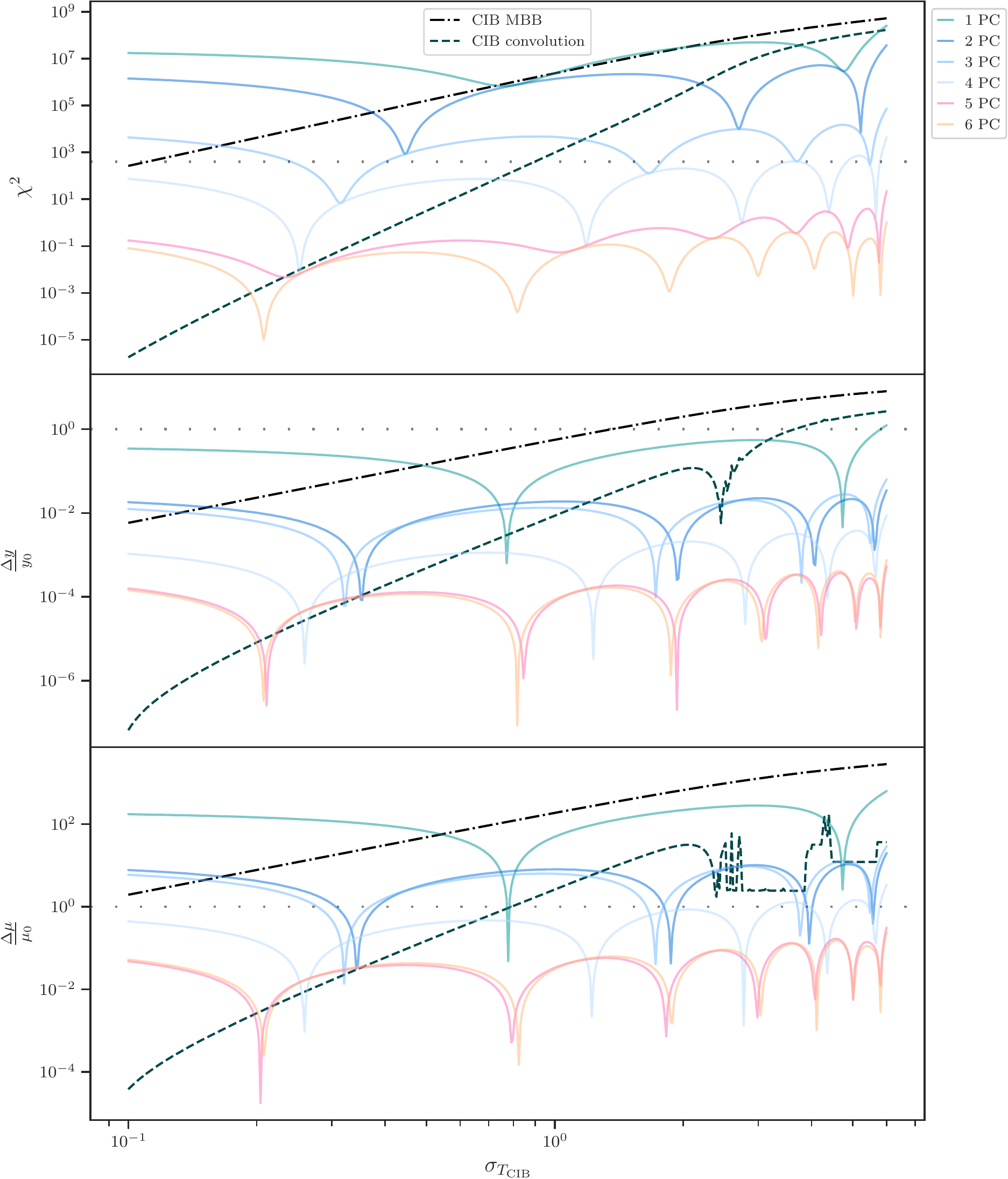}
	\caption{ Results from fitting the superposition of modified blackbody functions at different temperatures with 3 different procedures: a simple modified blackbody function, a PCA analysis with a varying number of CIB PC components, and a function derived from an MBB function convolved with a Gaussian of the same $\sigma_{T_{\textrm{CIB}}}$ as that of the reference broadened CIB SED. The dotted horizontal line in the first panel represents rougly $\chi^2$/degree of freedom = 1 (which happens at $\chi^2\approx 400$). These fits are performed only in the presence of the spectral distortions and in the absence of the other foregrounds.\label{fig:CIB_analysis}}
\end{figure*}

\begin{figure*}[ht!]
	\includegraphics[scale=0.94]{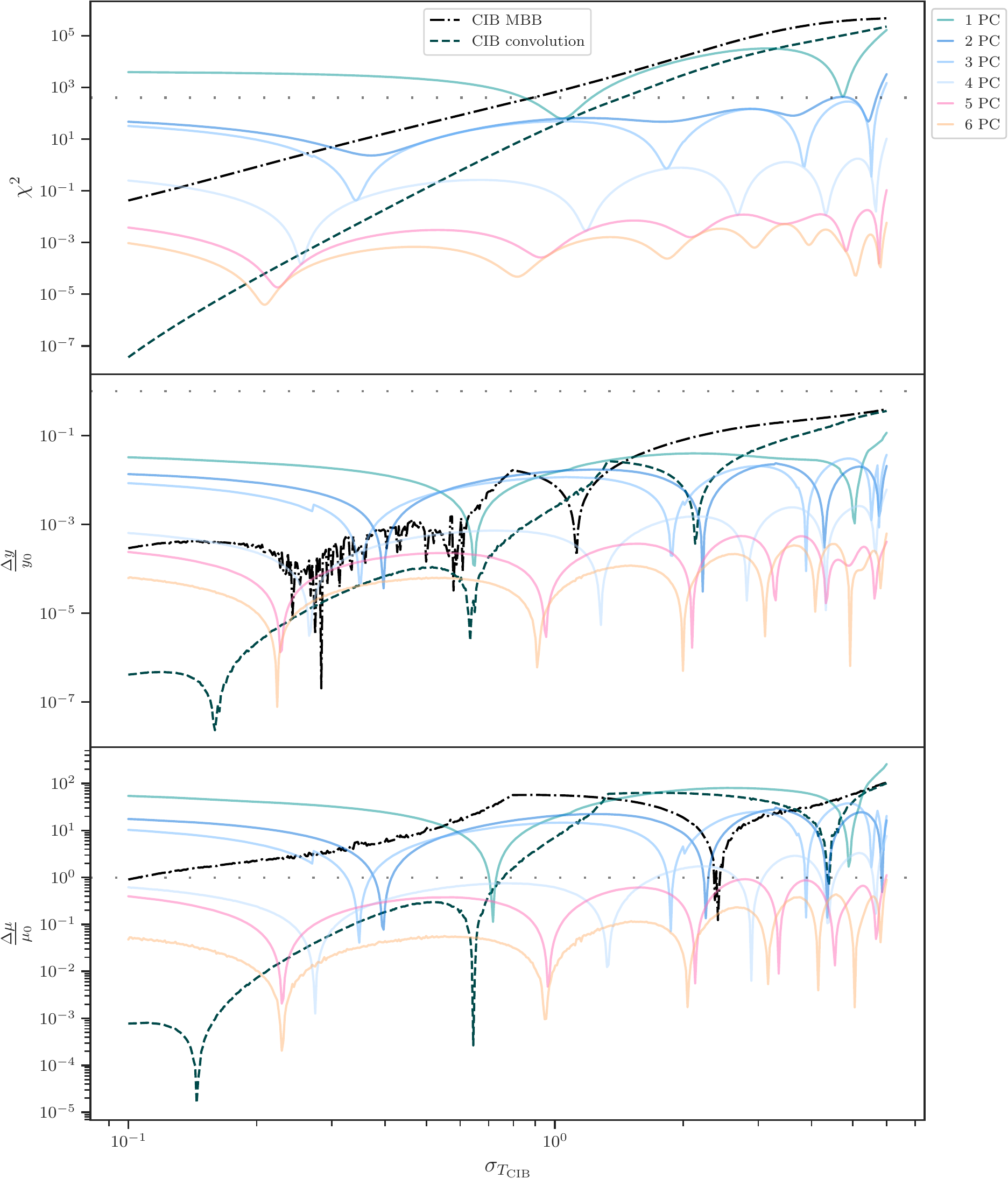}
	\caption{Results from fitting the superposition of modified blackbody functions at different temperatures with 3 different procedures: a simple modified blackbody function, a PCA analysis with a varying number of CIB PC components, and a function derived from an MBB function convolved with a Gaussian of the same $\sigma_{T_{\textrm{CIB}}}$ as that of the reference broadened CIB SED. The dotted horizontal line in the first panel represents roughly $\chi^2$/degree of freedom = 1 (which happens at $\chi^2\approx 400$). Unlike those in Fig. \ref{fig:CIB_analysis}, these fits include all the foregrounds. \label{fig:CIB_analysis_other_foregrounds}}
\end{figure*}

\section{CIB Analysis and Results}\label{sec:CIB}
Another important foreground for PIXIE is the cosmic infra-red background (CIB), consisting of far IR emission from all galaxy types integrated over a wide range of redshifts \citep{Hauser2001}.  Unlike ISM dust, which is predominantly in the Galactic plane, CIB is present everywhere on the sky, with a spectrum that varies on sub-degree scales.  This makes it even more difficult to model than ISM dust, and therefore makes it a potentially greater challenge for PIXIE.

Although the CIB is often modeled as an MBB, the spectrum is more varied than that. Planck has mapped the CIB over the full sky and found a spatial and spectral complexity that is beyond the scope of this paper to model \citep{Collaboration2014d, Collaboration2016f, Collaboration2016g, Lenz2019}.  Instead, we consider a linear combination of MBBs at various temperatures as a reference CIB spectrum, and compute the bias resulting from 3 approaches to modeling this departure from an MBB spectrum.  Our goal is not to choose one approach over another, but to use the disagreement between multiple reasonable approaches as an indication of how severe the modeling problem is.

\subsection{Reference CIB Spectrum}

To create a reference CIB spectrum, we compute a ``broadened'' MBB spectrum, i.e. an MBB averaged over a Gaussian distribution of temperatures (Fig. \ref{fig:CIB_broadening_example}).  This average is computed by drawing 1000 values for $T_{\textrm{CIB}}$ from a Gaussian with mean $\mu_{T_{\textrm{CIB}}}$=18.8K, and standard deviation $\sigma_{T_{\textrm{CIB}}}$.  We then examine the biases in $y$ and $\mu$ as a function of $\sigma_{T_{\textrm{CIB}}}$ using three different approaches:
\begin{enumerate}
	\item a simple modified blackbody function as given in Equation \ref{eq:CIB_MBB_shadow},
	\item a PCA analysis with a varying number of CIB PC components, and
	\item a function approximating an MBB function convolved with a Gaussian of the same $\sigma_{T_{\textrm{CIB}}}$ as that of the reference broadened CIB SED.
\end{enumerate}

The total reference function for the fit is calculated using:
\begin{equation}
\begin{split}
&\Delta I_{\nu}^{\textrm{reference}} =\Delta I_{\nu}^{\textrm{sync},\textrm{ broadened}}+  \Delta I_{\nu}^y + \Delta I_{\nu}^{\mu} +\\
&+\Delta I_{\nu}^{\Delta_T}+ \Delta I_{\nu}^{\textrm{dust,MBB}}+ \Delta I_{\nu}^{\textrm{CIB}} + \Delta I_{\nu}^{\textrm{FF}}
\end{split}
\end{equation}

The goodness of fit is given by $\chi^2 = \sum_{\nu}((\Delta I_{\nu}^{\textrm{total}}-\Delta I_{\nu}^{\textrm{reference}})/\sigma_{\nu}^{\textrm{PIXIE}})^2$, where the $\Delta I_{\nu}^{\textrm{total}}$ is the total intensity calculated by summing over all foreground components, with the CIB intensity defined according to each of the three methods used in the fit.

\subsection{Principal Component Analysis}

In analogy to the ISM dust PCA analysis above, we can use PCA to find the dimensionality of PIXIE spectral space spanned by broadened MBBs.  If a modest number of principal components can describe the CIB to sufficient precision, they could perhaps be marginalized over without losing too much information about $y$ and $\mu$.  

To obtain a set of spectra to input into the PCA, we generate 40000 broadened MBBs using the technique above.  Each MBB is broadened  by a different $\sigma_{T_{\textrm{CIB}}}$ drawn from a log-uniform distribution from 0.1-6K, then sampled at the 416 PIXIE frequencies.  

The intensity data is normalized by the PIXIE sensitivity before the principal component analysis is done.  As expected, the PC spectra show oscillatory behavior with PC$n$ having $n$ extrema (Fig. \ref{fig:CIB_principal_components}). PC1 accounts for 99.9\% of the variance, and keeping the first 5 components leaves $<10^{-10}$ of the variance behind (Table \ref{table:CIB_PC_variance}).

\subsection{A Broadened MBB}
Our third approach to CIB modeling is to explicitly broaden an MBB.  There is no analytical expression for the convolution of a Planck function with a Gaussian.

Appendix \ref{sec:gaussian_convolution} shows that the convolution of a function with a Gaussian can be approximated by an expansion in even derivatives multiplied by powers of $\sigma$ (Eq. \ref{eq:power_expansion}).

Applying this result to the modified blackbody function $\Delta I_{\nu}^{\textrm{CIB,s}}$ (Eq. \ref{eq:CIB_MBB_shadow}), we obtain the second order term:\footnote{The term is the same for CIB MBB with and without the CMB shadow component}
\begin{equation}
\begin{split}
\Delta I_{\nu}^{\textrm{CIB,s},(2)}=& \frac{\sigma_{\textrm{CIB}}^2}{2} \alpha_{\textrm{CIB}}\frac{\nu}{\nu_0}^{3+\beta_{CIB}}\\
&\times\left ( \frac{2 \exp( \frac{2h\nu}{k_b T_{\textrm{CIB}}} ) h^2\nu^2}{ \left(\exp( \frac{h\nu}{k_b T_{\textrm{CIB}}})-1\right)^3 k_b^2 T_{\textrm{CIB}}^4   }  \right .\\
& - \frac{ \exp( \frac{h\nu}{k_b T_{\textrm{CIB}}} ) h^2\nu^2}{ \left(\exp( \frac{h\nu}{k_b T_{\textrm{CIB}}})-1\right)^2 k_b^2 T_{\textrm{CIB}}^4   }\\
&-\left . \frac{ 2\exp( \frac{h\nu}{k_b T_{\textrm{CIB}}} ) h\nu}{ \left(\exp( \frac{h\nu}{k_b T_{\textrm{CIB}}})-1\right)^2 k_b T_{\textrm{CIB}}^3 } \right)
\end{split}
\end{equation}

Thus, up to the 2nd order term, we can write:
\begin{equation}
\Delta I_{\nu}^{\textrm{CIB,s, convolution}}(\sigma_{T_{\textrm{CIB}}} ) =   \Delta I_{\nu}^{\textrm{CIB,s}} + \Delta I_{\nu}^{\textrm{CIB,s},(2)}
\end{equation}

More terms could be included in the expansion, but we restrict it to second order after checking that higher-order terms do not diverge.

%%%%%%%%%%%%%%%%%%%%%%%%%%%%%%%%%%%%%%%%%%%%%%%%%%%

\subsection{CIB Analysis Results}

Figures \ref{fig:CIB_analysis} and \ref{fig:CIB_analysis_other_foregrounds} present the results of performing the fit using the three different methods, with and without other foregrounds.
 
Both the principal component analysis method and the convolution method perform better at fitting the broadened CIB than the modified blackbody function does. The $\chi^2$ and the errors on $y$ and $\mu$ are lower. 
For $\sigma_{T_{\textrm{CIB}}}<0.6$K, the convolution method produces fits that show the errors on the $y$ and $\mu$ distortions are still manageable (i.e., well below the fiducial values). At higher $\sigma_{T_{\textrm{CIB}}}$, the biases become larger than the fiducial values, indicating that a meaningful measurement will be difficult with PIXIE data alone.  

As we add more principal components, we are able to take the goodness of fit $\chi^2$ below levels of 1 per degree of freedom (for approximately 400 degrees of freedom) even at $\sigma_{T_{\textrm{CIB}}}$ of 6K. This is expected since as we increase the number of principal components, we match the model that generated the data better and better.  In a more realistic simulation with PIXIE measurement noise added, the $\chi^2$ would have a floor of 1 per degree of freedom. 

We conclude that using a model that did not generate the data will lead to large biases, especially as $\sigma_{T_{\textrm{CIB}}}$ increases beyond 0.6K.  Using a simple MBB to represent the CIB is clearly a poor approximation, though a modest number of principal components may be able to represent the CIB adequately. 

\cite{Chluba2017} also looked at the effect of averaging foregrounds over the beam, along the line-of-sight and across the sky. They proposed modeling the resulting average SEDs using a moment expansion of the original SED model. It would be interesting to know how many moments one would need to include for the modeling of CIB and other foregrounds for a mission like PIXIE, while checking for the impact on the detectability of CMB spectral distortions.

\begin{figure*}[t!]
	\includegraphics[scale=1]{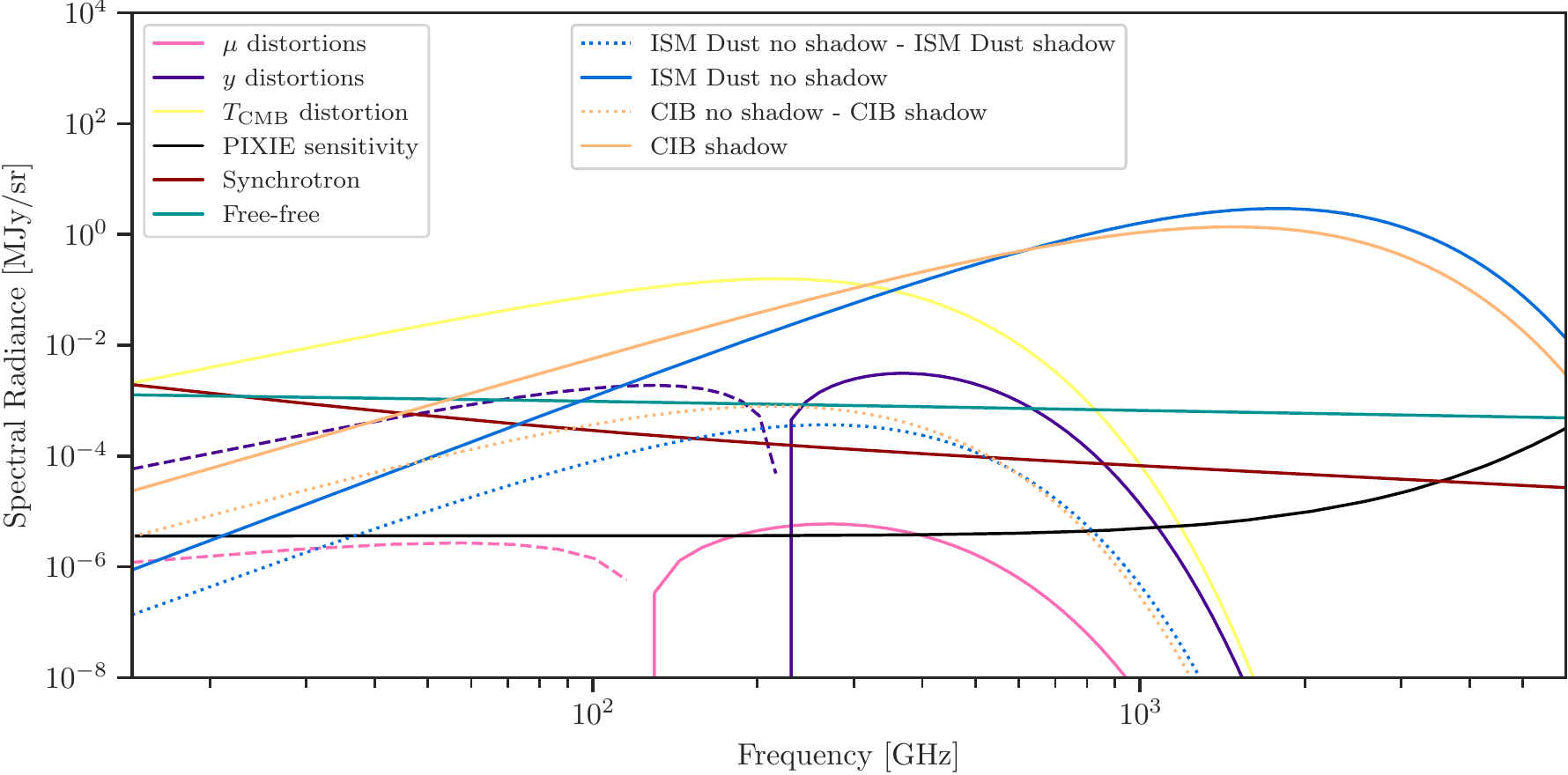}
	\caption{The blue and orange dotted lines represent the amplitude of the CMB absorption caused by the Galactic and extragalactic dust. \label{fig:shadow_foreground}}
\end{figure*}

\section{CMB Shadows}\label{sec:CMB_shadows}
% Interstellar dust extinction has long been neglected at microwave frequencies, but the increasing sensitivity of Cosmic Microwave Background observations (at the part in $10^8$ level) requires that we pay attention to subtle effects.  Scattering of the CMB monopole has no net effect, but dust absorption of the monopole removes energy from the CMB spectrum and emits it at much higher frequencies. \cite{Nashimoto2020} showed that interstellar absorption (they call it \emph{CMB shadows}) is important for CMB polarization and temperature anisotropy. Here we study the impact of CMB shadows (both ISM dust and CIB) on the detectability of $y$ and $\mu$ (Figure \ref{fig:shadow_foreground}).

Interstellar dust extinction has long been neglected at microwave frequencies, but the increasing sensitivity of Cosmic Microwave Background observations (at a part in $10^8$ level) requires that we pay attention to subtle effects. \cite{Nashimoto2020} showed that interstellar absorption (they called it \emph{CMB shadows}) is important for CMB polarization and temperature anisotropy. While scattering of the CMB monopole has no net effect, they claimed dust absorption of the monopole removes energy from the CMB spectrum and emits it at much higher frequencies. There is some question whether this interpretation is correct. \cite{Wright1991} found that if the energy levels are low enough where the harmonic approximation still applies, stimulated emission from dust grains cancels the absorption of the CMB. In future work, we aim to explore the regimes in which this approximation breaks down, and whether or not it matters for CMB science. For this work, taking \cite{Nashimoto2020}'s argument as valid, we study the impact of CMB shadows (both ISM dust and CIB) on the detectability of $y$ and $\mu$ (Figure \ref{fig:shadow_foreground}).

\begin{figure*}[t!]
    \hspace{-7mm}
	\includegraphics[scale=0.27]{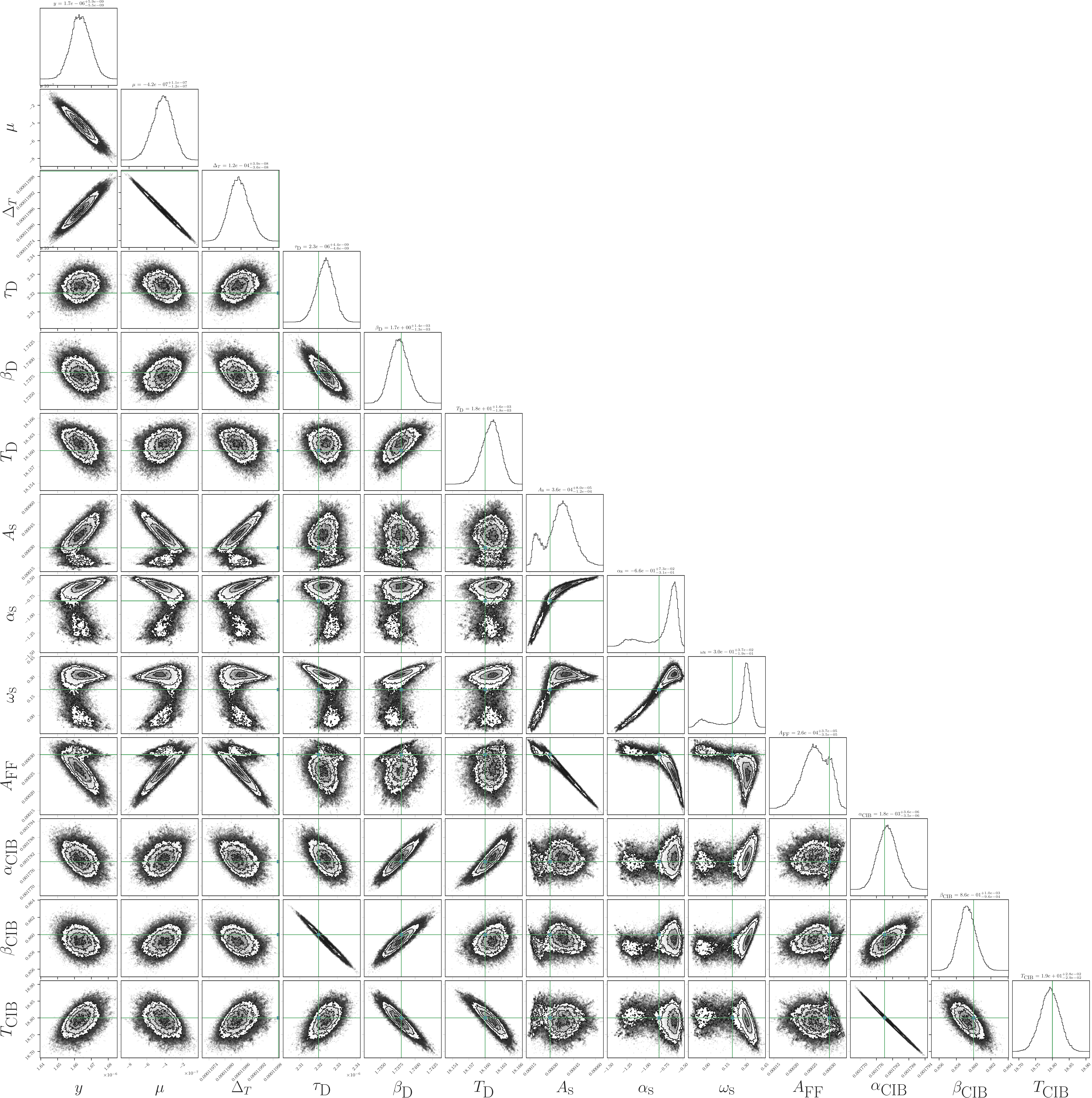}
	\caption{The posterior resulting from trying to model a reference function created with CMB shadows coming from the ISM dust, using an ISM dust model with no CMB shadow. \label{fig:shadow_ISM}}
\end{figure*}

For the case of ISM dust, we calculate a reference intensity using an MBB with the CMB dust shadows (Eq.~\ref{eq:MBB_ISM_dust_shadow}). For the fit, intensity from the ISM is calculated using an MBB without shadows. The other foregrounds are added to both the fit and the reference. An MCMC is run using the same sampler parameters as described in \S \ref{sec:MCMC_foreground}. Figure \ref{fig:shadow_ISM} and Table \ref{table:shadow_ISM} show the posterior resulting from trying to model a reference function with dust shadows, using an ISM dust model with no CMB shadow. The median of the $y$ distortion parameter shifts from 1.70E-6 to 1.66E-6, a 2.4$\%$ change. The bias in $\mu$ is an order of magnitude higher than the fiducial value. This simple test indicates that the CMB shadow caused by ISM dust must be included in the modeling.

\begin{figure*}[t!]
    \hspace{-7mm}
	\includegraphics[scale=0.27]{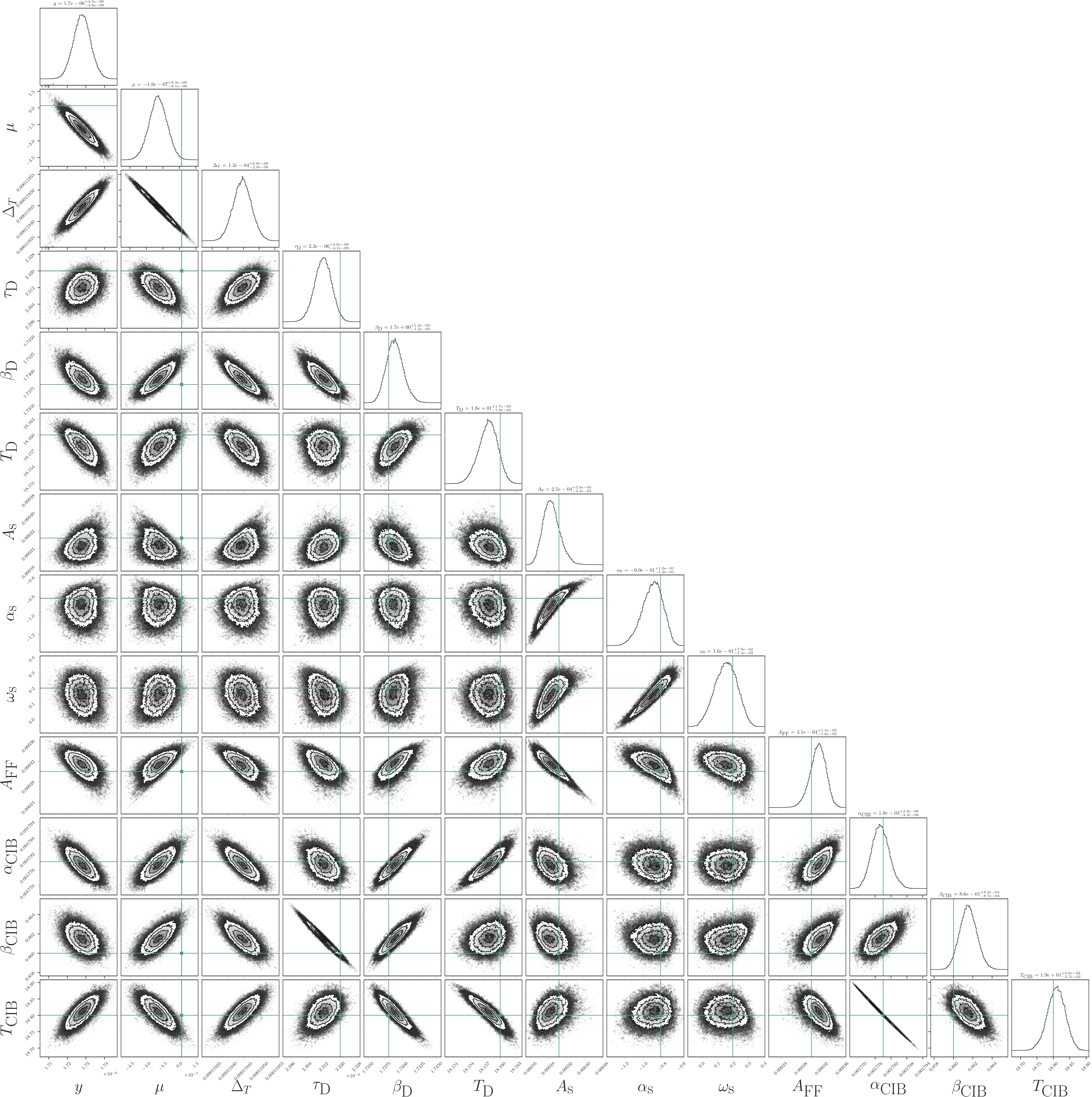}
	\caption{The posterior resulting from trying to model a reference function created with CMB shadows coming from the CIB, using a CIB model with no CMB shadow\label{fig:shadow_CIB}}
\end{figure*}

We apply the same procedure to exploring the CMB shadows caused by the dust in other galaxies (CIB). We use Eq. \ref{eq:CIB_MBB_shadow} to generate the reference intensity that includes the impact of the shadows, and Eq. \ref{eq:CIB_MBB_no_shadow} to generate the fit intensity of an MBB with no shadows. We run an MCMC, and obtain the results shown in Figure \ref{fig:shadow_CIB} and Table \ref{table:shadow_CIB}.  The median of the $y$ distortion parameter shifts from 1.70E-6 to 1.73E-6, a 1.7\% change. Here as well, the bias in $\mu$ is an order of magnitude higher than the fiducial value.
As a result, we conclude that the CMB monopole extinction caused by the CIB should be accounted for in the analysis.

\cite{Nashimoto2020} compared the CMB shadows produced by the other foregrounds (synchrotron, free-free) to that of dust, and found them to be negligible. We expect that to be the case for spectral distortions as well.
\begin{table*}[ht!]
\setlength{\tabcolsep}{1.5pt}
\centering
\footnotesize
\begin{tabular}{rrrrrrrrrrrrrr}
\hline
Param Name & $y$&$\mu$&$\Delta_T$&$\tau_{\textrm{D}}$&$\beta_{\textrm{D}}$&$T_{\textrm{D}}$&$A_{\textrm{s}}$&$\alpha_{\textrm{s}}$&$\omega_{\textrm{s}}$&$A_{\textrm{FF}}$&$\alpha_{\textrm{CIB}}$&$\beta_{\textrm{CIB}}$&$T_{\textrm{CIB}}$ \\ \hline
Units & -&-&[K]&-&-&[K]&[MJy/sr]&-&-&[MJy/sr]&[MJy/sr]&-&[K] \\ \hline
Fiducial Val. & 1.70e-06&2.00e-08&1.20e-04&2.32e-06&1.74e+00&1.82e+01&2.88e-04&-8.20e-01&2.00e-01&3.00e-04&1.78e-03&8.60e-01&1.88e+01 \\ \hline
Post. Median & 1.66e-06&-4.20e-07&1.20e-04&2.32e-06&1.74e+00&1.82e+01&3.61e-04&-6.62e-01&2.98e-01&2.64e-04&1.78e-03&8.59e-01&1.88e+01 \\ \hline
$\sigma_{+}$ & 5.91e-09&1.09e-07&3.95e-08&4.43e-09&1.42e-03&1.59e-03&8.02e-05&7.30e-02&3.73e-02&3.71e-05&3.66e-06&1.01e-03&2.86e-02 \\ \hline
$\sigma_{-}$ & 5.58e-09&1.18e-07&3.59e-08&4.63e-09&1.32e-03&1.81e-03&1.21e-04&3.15e-01&1.94e-01&3.54e-05&3.52e-06&9.66e-04&2.90e-02 \\ \hline
\end{tabular}
\caption{The results of the MCMC for the configuration testing the importance of modeling the ISM dust shadow, with CIB, synchrotron, and free-free foregrounds. \label{table:shadow_ISM}$ \sigma_{+}$ represents the difference between the 84th and 50th quantile of the posterior, and $\sigma{-}$ the one between the 50th and 16th. We observe that failing to include the ISM dust shadow results in a small bias in $y$ and a very large bias in $\mu$.}
\end{table*}

\begin{table*}[ht]
\setlength{\tabcolsep}{1.5pt}
\centering
\footnotesize
\begin{tabular}{rrrrrrrrrrrrrr}
\hline
Param Name & $y$&$\mu$&$\Delta_T$&$\tau_{\textrm{D}}$&$\beta_{\textrm{D}}$&$T_{\textrm{D}}$&$A_{\textrm{s}}$&$\alpha_{\textrm{s}}$&$\omega_{\textrm{s}}$&$A_{\textrm{FF}}$&$\alpha_{\textrm{CIB}}$&$\beta_{\textrm{CIB}}$&$T_{\textrm{CIB}}$ \\ \hline
Units & -&-&[K]&-&-&[K]&[MJy/sr]&-&-&[MJy/sr]&[MJy/sr]&-&[K] \\ \hline
Fiducial Val. & 1.70e-06&2.00e-08&1.20e-04&2.32e-06&1.74e+00&1.82e+01&2.88e-04&-8.20e-01&2.00e-01&3.00e-04&1.78e-03&8.60e-01&1.88e+01 \\ \hline
Post. Median & 1.73e-06&-1.94e-07&1.19e-04&2.31e-06&1.74e+00&1.82e+01&2.52e-04&-9.00e-01&1.58e-01&3.13e-04&1.78e-03&8.62e-01&1.88e+01 \\ \hline
$\sigma_{+}$ & 4.57e-09&8.25e-08&2.90e-08&4.02e-09&1.26e-03&1.72e-03&3.97e-05&1.03e-01&7.07e-02&1.45e-05&3.39e-06&8.99e-04&2.66e-02 \\ \hline
$\sigma_{-}$ & 4.62e-09&8.11e-08&2.95e-08&4.12e-09&1.18e-03&1.90e-03&3.45e-05&1.17e-01&7.47e-02&1.61e-05&3.20e-06&8.71e-04&2.75e-02 \\ \hline
\end{tabular}
\caption{The results of the MCMC for the configuration testing the effect of not modeling the CIB shadow, with ISM dust, synchrotron and free-free foregrounds. \label{table:shadow_CIB}$ \sigma_{+}$ represents the difference between the 84th and 50th quantile of the posterior, and $\sigma{-}$ the one between the 50th and 16th. We observe that failing to include the CIB shadow results in a small bias in $y$ and a very large bias in $\mu$.}
\end{table*}

\section{Conclusions}\label{sec:conclusions}

The primary goal of this work is to quantify the variability of dust emission in the PIXIE frequency range, and estimate the sensitivity of PIXIE to $y$ and $\mu$ distortions after marginalizing over dust emission.  In order for this estimate to be meaningful, it is \emph{not} required that the dust model be correct, merely that it reflect the range of possible variations in the size distribution and composition for interstellar dust.  A model with reasonable constituents and a plausible range of grain-size distributions, constrained by well motivated physical priors, provides at least a \emph{lower bound} on the variation in the emission spectrum.   

Previous work used modified blackbody (MBB) models to represent both interstellar and extragalactic dust \citep{Abitbol2017}. In this work, we use a more general dust model that takes into account the abundance, composition, size, and temperature distribution of the grains, as well as the interstellar radiation field. This is done in the context of the Draine models \citep{Draine1984, Li2001, Weingartner2001a}, constrained to be consistent with the observed spectral dependence of reddening in the visible-near IR, and the $R(V)-\beta$ relation \citep{Schlafly2016}.  Our previous work \citep{Zelko2020} generated samples from the posterior on dust composition and size distribution used in this work.  

For each sample from the size distribution space we generate the dust emission spectra. These spectra feed into our PCA to determine what directions in the PIXIE spectral space correspond to dust, and therefore must be marginalized over. We perform an MCMC and Fisher information matrix analysis to determine how much the sensitivity to  $y$ and $\mu$ degrades as a function of the complexity of the dust model. We observe that  most of the PIXIE noise-weighted variance in spectral space is contained in the first two principal components, and adding up to 20 principal components (i.e. marginalizing over 20 free parameters describing dust) increases the standard deviation for the $y$ and $\mu$ distortion parameters by up to an order of magnitude. Thus, complicating the dust model and including up to 20 principal components still keeps the detectability of $y$ and $\mu$ plausible. When including other foregrounds like synchrotron radiation, the cosmic infrared background, and free-free emission, as well as the spectral distortion from the blackbody temperature deviation, the $y$ and $\mu$ expected standard deviation increases. The $y$ detectability remains, while that of $\mu$ depends on how well ancillary information can be used to constrain the foregrounds, and the strength of the expected $\mu$ distortion. Unless other datasets are provided to constrain some of the foregrounds, the $\mu$ distortion will not be detectable in the range sometimes considered for the standard $\Lambda$CDM of $2\times10^{-8}$, but could potentially remain detectable at higher values. Even a signal 1000 times stronger than that would not have been detectable by FIRAS, so there is a lot of new parameter space available.

It is inevitable that any dust model will be incomplete, giving rise to a model discrepancy error. We make a simple estimate of this error by generating dust spectra based on the Draine models, and fitting them with an MBB. The bias this introduces is orders of magnitude larger than the fiducial values for $y$ and $\mu$. For CMB spectral distortions to be detectable, the ISM dust model must contain degrees of freedom that cover the relevant spectral subspace.
 
The CIB presents a significant challenge, given its spectral and spatial variation over the full sky.  We create a mock CIB SED from a superposition of modified blackbodies over a range of temperatures. The effect of attempting to fit this using a simple MBB, a set of PCA spectra, and a broadened MBB is calculated. A simple MBB is a poor approximation, and the broadened MBB is somewhat better. However, using a few principal components may be able to capture the CIB adequately, at least in the restricted model space we explore. Again, using a model that did not generate the data will lead to large biases, especially as the distribution of MBB temperatures making up the CIB, $\sigma_{T_{\textrm{CIB}}}$, broadens beyond about 0.5K (and in reality it is surely larger).

Finally, we examine the impact of the absorption of the CMB monopole by the ISM dust and extragalactic dust, as described in \cite{Nashimoto2020}. Failing to model this effect results in a small bias in $y$ and a catastrophic bias in $\mu$. We conclude that the CMB shadow due to both ISM dust and dust in other galaxies must be considered in the future CMB spectral distortion measurements in order to achieve their desired sensitivity.

\paragraph{Public Supporting Data Set} The data set containing the dust spectral energy distribution data from \cite{Zelko2020}  and the current paper is available publicly at the Harvard Dataverse repository \citep{Zelko}\footnote{\url{https://dataverse.harvard.edu/dataset.xhtml?persistentId=doi:10.7910/DVN/DJAAIG}}.

\paragraph{Acknowledgments}
We acknowledge helpful conversations with Ana Bonaca, Blakesley Burkhart, Erin Cram, Tansu Daylan, Bruce Draine, Cora Dvorkin,  Daniel Eisenstein, Daniel Foreman-Mackey, J. Colin Hill, John Kovac, Albert Lee, Karin \"{O}berg, Stephen Portillo, Eddie Schlafly, Zachary Slepian, Josh Speagle, David Spergel, Justina Yang, Jun Yin and Catherine Zucker. IZ is supported by the Harvard College Observatory. DF is partially supported by  National Science Foundation grant AST-1614941, ``Exploring the Galaxy: 3-Dimensional Structure and Stellar Streams.'' 
This research made use of the NASA Astrophysics Data System's Bibliographic Services (ADS), the Odyssey Cluster at Harvard University, the color blindness palette by Martin Krzywinski and Jonathan Corum\footnote{\url{http://mkweb.bcgsc.ca/biovis2012/color-blindness-palette.png}}, and the Color Vision Deficiency PDF Viewer by Marie Chatfield \footnote{\url{https://mariechatfield.com/simple-pdf-viewer/}}.

\software{ptemcee \citep{Vousden2016}, NumPy \citep{VanderWalt2011}, Matplotlib \citep{Hunter2007}, pandas \cite{mckinney-proc-scipy-2010}, scikit-learn \citep{Pedregosa2012}, IPython \citep{Perez2007}, Python \citep{Millman2011, Oliphant2007}}

\appendix
\section{Fisher Information Matrix Derivatives Calculation}\label{sec:derivative_calculation}
This section contains the analytical forms of the function derivatives used for the Fisher information matrix calculations in \S \ref{sec:ISM_dust_forecasting_methods}.

\paragraph{$\mu$ distortion}

\begin{equation}
\frac{\partial \Delta I_{\nu}^{\mu}}{\partial \mu} = I_0\frac{x^4e^x}{(e^x-1)^2}\bigg[\frac{1}{\beta}-\frac{1}{x}\bigg]
\end{equation}

\paragraph{y Distortion}
\begin{equation}
\frac{\partial\Delta I_{\nu}^y}{\partial y} = I_0 \frac{x^4e^x}{(e^x-1)^2}\bigg[x \coth\bigg(\frac{x}{2}\bigg) -4 \bigg] 
\end{equation}
\paragraph{Blackbody Temperature Distortion}
\begin{equation}
\pdv{\Delta I_{\nu}^{\Delta_T}}{\Delta_T} = I_0 \frac{x^4 e^x}{(e^x-1)^2} 
\end{equation}
\paragraph{ISM Dust Principal Components}

\begin{equation}
\pdv{\Delta I^\textrm{dust PC}_{\nu}}{\textrm{WC}_i} = \textrm{PC}_i\times \sigma^\textrm{PIXIE}_{\nu}
\end{equation}

\paragraph{Synchrotron}

\begin{equation}
\pdv{\Delta I^{\textrm{s}}_{\nu}}{A_{\textrm{s}}} =  \bigg (\frac{\nu}{\nu_0} \bigg )^{\alpha_\textrm{s}}\big(1+\frac{1}{2}\omega_{\textrm{s}} \ln^2{(\frac{\nu}{\nu_0})}\big) 
\end{equation}

\begin{equation}
\pdv{\Delta I^{\textrm{s}}_{\nu}}{\alpha_\textrm{s}} = A_{\textrm{s}}  \bigg (\frac{\nu}{\nu_0} \bigg )^{\alpha_\textrm{s}}\big(1+\frac{1}{2}\omega_{\textrm{s}} \ln^2{(\frac{\nu}{\nu_0})}\big)
\ln{(\frac{\nu}{\nu_0})}
\label{eq:ap_power_sinc}
 \end{equation}

\begin{equation}
\pdv{\Delta I^{\textrm{s}}_{\nu}}{\omega_{\textrm{s}}}  = \frac{A_{\textrm{s}}}{2}  \bigg (\frac{\nu}{\nu_0} \bigg )^{\alpha_\textrm{s}} \ln^2{(\frac{\nu}{\nu_0})} 
\end{equation}

\paragraph{Cosmic Infrared Background}

\begin{equation}
\pdv{\Delta I^{\textrm{CIB}}_{\nu}}{\alpha_{\textrm{CIB}}} = \left(\frac{\nu}{\nu_0}\right)^{\beta_{\textrm{CIB}}} \left( \frac{(\nu/\nu_0)^3}{e^{h\nu/k/T_{\textrm{CIB}} }-1} - \frac{(\nu/\nu_0)^3}{e^{h\nu/k/T_0 }-1} \right)
\end{equation}

\begin{equation}
\pdv{\Delta I^{\textrm{CIB}}_{\nu}}{\beta_{\textrm{CIB}}} = \beta_{\textrm{CIB}} \alpha_{\textrm{CIB}} \left(\frac{\nu}{\nu_0}\right)^{\beta_{\textrm{CIB}}} \left( \frac{(\nu/\nu_0)^3}{e^{h\nu/k/T_{\textrm{CIB}} }-1} - \frac{(\nu/\nu_0)^3}{e^{h\nu/k/T_0 }-1} \right)\ln{(\frac{\nu}{\nu_0})}
\label{eq:ap_power_CIB}
\end{equation}

\begin{equation}
\pdv{\Delta I^{\textrm{CIB}}_{\nu}}{T_{\textrm{CIB}}} = \alpha_{\textrm{CIB}} \left(\frac{\nu}{\nu_0}\right)^{\beta_{\textrm{CIB}}} \left( \frac{(\nu/\nu_0)^3 \frac{h\nu}{k T_{\textrm{CIB}}^2}e^{h\nu/k/T_{\textrm{CIB}} }}{\left(e^{h\nu/k/T_{\textrm{CIB}} }-1\right)^2}\right)
\end{equation}

\paragraph{Free-free Emission}

\begin{equation}
\pdv{\Delta I^{\textrm{FF}}_{\nu}}{A_{\textrm{FF}}} = 1+\ln{[1+(\frac{\nu_{\textrm{ff}}}{\nu})^{\sqrt{3}/\pi}]}
\end{equation}

\section{Analytical Approximation to Gaussian Convolution}\label{sec:gaussian_convolution}
In this section we derive the analytical approximation to convolving a function $f$ with a Gaussian.
We use a normal Gaussian because we do not want to scale the function up or down, but rather to  just"smooth it over". For our purposes, the mean of the Gaussian has to be zero, otherwise the convolution would shift the original function as well. The Gaussian is approximated using two $\delta$ functions\footnote{If we only pick one $\delta$ function, it will have a zero standard deviation. So, we use two $\delta$ functions.}, in such a way that the resulting function has the same mean and standard deviation of the Gaussian. Because the amplitude of a delta function when integrated over an entire domain is one, we scale it by a factor of 2: 
\begin{equation}
g(T,0,\sigma)\approx \frac{\delta(T-\Delta T)+\delta(T+\Delta T)}{2}=h(T,0,\Delta T)
\end{equation}
We want to know what should be the offset $\Delta T$ such that the function $h$  has the same standard deviation as the Gaussian. Thus, the offset $\Delta T$ can be calculated as a function of the desired $\sigma$ in the following manner:
\begin{equation}
\begin{split}
\sigma & = \sqrt{\int{(T -\mu)^2 h(T,\mu,\Delta T)}\dd T}   = \sqrt{\int{T^2 h(T,0,\Delta T)\dd T}} \\
& = \sqrt{\int{T^2 \frac{\delta(T-\Delta T)+\delta(T+\Delta T)}{2} \dd T}}\\
& = \left (\frac{1}{2}\int{T^2\delta(T-\Delta T)\dd T} + \frac{1}{2}\int{T^2\delta(T+\Delta T)\dd T} \right)^{\frac{1}{2}}\\
& = (\frac{1}{2} [ (\Delta T )^2 + (\Delta T)^2] )^{\frac{1}{2}}\\
& = \Delta T 
\end{split}
\end{equation}
This proves that $\Delta T$ should be equal to $\sigma$. As a result, the approximation to the Gaussian will be given by:
\begin{equation}
g(T,0,\sigma) = g(T,\sigma) \approx \frac{\delta(T-\sigma)+\delta(T+\sigma)}{2}
\end{equation}

Now, the convolution of a function $f(T)$ with the Gaussian approximation will be given by: 
\begin{equation}\label{eq:convolution}
\begin{split}
f \circledast g(\sigma))(\tau) &= \int{f(T)g(\tau-T,\sigma)\dd T}  \\
&\approx  \int{f(T) \frac{\delta(\tau-T-\sigma)+\delta(\tau-T+\sigma)}{2} \dd T} \\
& = \frac{1}{2} \int{f(T) \delta(\tau-T-\sigma) \dd T} + \frac{1}{2} \int{f(T) \delta(\tau-T+\sigma) \dd T}\\
& =  \frac{1}{2} \int{f(T) \delta(T+\sigma-\tau) \dd T} + \frac{1}{2} \int{f(T) \delta(T-\sigma-\tau) \dd T}\\
& = \frac{1}{2}(f(\tau-\sigma)+ f(\tau+\sigma)),
\end{split}
\end{equation}
where we make use of the relation $\delta(ax) = \frac{1}{|a|}\delta(x)$.

We represent the function $f(T)$ as a Taylor Series:
\begin{equation}\label{eq:taylor_expansion}
\begin{split}
&f(T-\sigma) = f(T) + \sum_{n=1}^{n=\infty} \frac{f^{(n)}(T)}{n!}(-\sigma)^{n}\\
&f(T+\sigma) = f(T) + \sum_{n=1}^{n=\infty} \frac{f^{(n)}(T)}{n!}\sigma^{n}
\end{split}
\end{equation}

Plugging Eq. \ref{eq:taylor_expansion} into Eq. \ref{eq:convolution}, we obtain:

\begin{equation}\label{eq:power_expansion}
\begin{split}
f \circledast g(\sigma))(T) -f(T)& = \frac{1}{2}\left( f(T-\sigma)+ 
f(T+\sigma)\right)-f(T) \\
&= \frac{1}{2}f''(T)\sigma^2 + \frac{1}{4!}f\textsuperscript{(4)}(T)\sigma^4 +\cdots\\
&= \sum_{n=1}^{\infty}\frac{1}{(2n)!}f\textsuperscript{(2n)}\sigma^{2n}
\end{split}
\end{equation}

Thus, the convolution of a function with a Gaussian can be approximated by an expansion in even derivatives multiplied by the power terms of $\sigma$.

\bibliography{CMB.bib}
\end{document}